\documentstyle[preprint,eqsecnum,aps,here,psfig]{revtex}
\newcommand{\twocol}{\narrowtext}
\newcommand{\onecol}{\widetext}

\newcommand{\bm}[1]{\mbox{\boldmath$#1$}}
\newcommand{\ben}{\begin{eqnarray}}
\newcommand{\een}{\end{eqnarray}}
\begin{document}
\draft
\title{PHOTOABSORPTION ON A NUCLEON \\
IN THE D$_{13}$ RESONANCE ENERGY REGION}
\author{
Kazuyuki O{\small CHI}
\thanks{
e-mail:   kazu@theo.phys.sci.hiroshima-u.ac.jp} 
,
Michihiro H{\small IRATA}
}
\address{
Department of Physics, Hiroshima University, Higashi-Hiroshima 739, Japan}
\author{
and \\
Takashi T{\small AKAKI}}
\address{
Onomichi Junior College, Onomichi 722, Japan}
\maketitle
\begin{abstract}
We present a simple model for the
$\gamma N \rightarrow \pi \pi N$ reaction
which reproduces the cross sections of the $\pi^{+} \pi^{-}p$,
$\pi^{+} \pi^{-}n$, $\pi^{+} \pi^{0}n$ and $\pi^{-} \pi^{0}p$
channels over the range of the energies 0.41$-$0.85 GeV.
We use the dynamical model for the resonances,
$\Delta$(1232), $N^{*}$(1520) and $\rho$-meson.
The total photoabsorption off a nucleon is also discussed.
\end{abstract}
\twocol
\section{Introduction}
The total nuclear-photoabsorption cross sections through
the third resonance (0.2 $-$ 1.2 GeV photon laboratory energy)
have been measured for a wide set of nuclei to study the behavior
of baryon resonances\cite{Nucl1,Nucl2,Nucl3,Nucl4,Nucl5} in nuclear matter.
These measurements show an interesting result.
In the first resonance region, the P$_{33}$ resonance
($l=1$, $J=3/2$, $I=3/2$) is slightly distorted.
In the second and third resonance region, on the other hand, the total
cross section is largely suppressed and the resonances such as
D$_{13}$($l=2$, $J=3/2$, $I=1/2$) and F$_{15}$($l=3$, $J=5/2$, $I=1/2$)
disappear in the excitation function.
Apparently, these resonances in the fundamental process, i.e.,
pion photoproduction off a nucleon,
are strongly modified by nuclear medium effects.
There have been several theoretical
works\cite{Phenomenan1,Phenomenan2,Phenomenan3} to explain the disappearance
of the resonances.
In some phenomenological analyses\cite{Phenomenan1,Phenomenan2},
very large collision broadening have been assumed to fit the data.
In the other work\cite{Phenomenan3}, however, it has been claimed that such
significantly increasing resonance widths were hardly justified.
The puzzle regarding the mechanism of resonance disappearance still
remains unresolved.

In the D$_{13}$ and F$_{15}$ resonance regions, the double pion 
photoproduction ($\gamma N \rightarrow \pi \pi N$)
is important on pion photoproduction in addition to the single pion
photoproduction ($\gamma N \rightarrow \pi N$).
In those processes, the $N^{*}(1520)$($J^{P}= 3/2^{-},I= 1/2 $)
and the $N^{*}(1680)$($J^{P}=5/2^{+},I=1/2 $) resonances play a 
significant role as an intermediate state.
In order to investigate the unknown mechanism that has caused 
resonances damping, one needs precise information about those
fundamental pion photoproduction processes.
%

The $\gamma N \rightarrow \pi N$ reaction
has been studied experimentally in the past\cite{singlepi-exp}.
Moorhouse {\it et al}.\cite{Moorhouse} and Arai {\it et al}.\cite{Fujii}
analyzed the $\gamma N \rightarrow \pi N$ reaction
data from the first through the third resonance region.
They made a partial-wave analysis of the processes
$\gamma p \rightarrow \pi^{+}n$, $\gamma p \rightarrow \pi^{0}p$,
and $\gamma n \rightarrow \pi^{-}p$.
The imaginary parts of the amplitudes were parameterized with $K$-matrices
written as a sum of factorizable poles.
The real parts of the amplitudes are calculated from the imaginary parts
through the fixed-$t$ dispersion relations.
Furthermore, Arndt {\it et al}.\cite{Arndt} made an energy-dependent
partial-wave analysis on the data for the processes
$\gamma p \rightarrow \pi^{+} n$, $\gamma p \rightarrow \pi^{0} p$,
$\gamma n \rightarrow \pi^{-} p$, $\gamma n \rightarrow \pi^{0} n$
and the inverse reaction $\pi^{-} p \rightarrow n \gamma $.
From their analyses, D$_{13}$ and F$_{15}$ resonances have been found
to be important in the $\gamma N \rightarrow \pi N$ reaction
in addition to the P$_{33}$ resonance.

The $\gamma p \rightarrow \pi ^{+} \pi ^{-} p$
reaction cross section has been measured
as a function of the photon energy (0.3 $-$ 5.8 GeV)
by ABBHHM collaboration in 1968\cite{doublepi-exp1}.
Recently, new improvement in experimental techniques make them possible
to study the $\gamma p \rightarrow \pi^{+} \pi^{-}p$ more accurately
and to observe other isospin channels, i.e.,
$\gamma p \rightarrow \pi^{0} \pi^{+} n$ and
$\gamma p \rightarrow \pi^{0} \pi^{0} p$.
These new data have been obtained at Mainz for photon energy ranging 0.45
$-$ 0.8 GeV\cite{doublepi-exp2}.
Theoretically, the $\gamma N \rightarrow \pi \pi N$
reactions were studied by Tejedor {\it et al}.\cite{Oset}
and Murphy {\it et al}.\cite{Laget}. 
Their studies show that their models can reproduce the
$\gamma p \rightarrow \pi^{+} \pi^{-}p$ 
reaction cross sections fairly well, 
but fail to explain the
$\gamma p \rightarrow \pi^{+} \pi^{0} n$ cross section.
Tejedor {\it et al}., furthermore, calculated the cross sections
for the neutron target,
but the results are not in good agreement with the data\cite{PIA,CAR}.
In order to investigate the total nuclear photoabsorption cross section,
their models should be improved.

In this paper we propose a modified model
for the $\gamma N \rightarrow \pi \pi N$ reaction,
taking into account both $\gamma N \rightarrow \pi N$ 
and $\gamma N \rightarrow \pi \pi N$ reactions consistently and
treating the $\rho$-meson propagation carefully.
The self-energy of the $\rho$-meson is calculated assuming the
$\rho \pi \pi$ form factor.
We focus on the photon energy range measured at Mainz\cite{doublepi-exp2}.
So, we include only the $\Delta (1232)$ and $N^{*}(1520)$
as intermediate baryon resonance states in our model. 
The $N^{*}(1520)$ will be treated carefully since it 
contributes importantly to the $\gamma N \rightarrow \pi N$ and 
$\gamma N \rightarrow \pi \pi N$ reactions. 
The dominant $\Delta$ Kroll-Ruderman term is constructed from the
finite-ranged form factor of the $\pi N \Delta$ by requiring the
gauge invariance.

This paper is organized as follows.
In Sec.\ref{sec:Nstar} we discuss how to describe the
D$_{13}$ amplitude and the $N^{*}(1520)$ propagator and furthermore
how to determine strong coupling constants such as
$\pi N N^{*}(1520)$, $\pi \Delta N^{*}(1520)$
and $\rho N N^{*}(1520)$ in detail.
In Sec.\ref{sec:photoncoupling} we discuss how to obtain the
$\gamma N N^{*}(1520)$ coupling constant.
In Sec.\ref{sec:2p}
we present our model of the
$\gamma N \rightarrow \pi \pi N$ reaction, which is based on the formalism
given in Secs.\ref{sec:Nstar} and \ref{sec:photoncoupling}.
In Sec.\ref{sec:discussion}, we show our predictions of the
$\gamma N \rightarrow \pi \pi N$ cross sections and the total photoabsorption
cross sections and then we discuss the numerical results.
Finally we give concluding remarks in Sec.\ref{sec:conclusion}.
\section{$N^{*}$(1520) resonance}
\label{sec:Nstar}
The $N^{*}(1520)$ resonance can decay into both the $\pi N$
and $\pi \pi N$ channels.
The branching fractions to $\pi N$ and $\pi \pi N$ are
50$\sim$60\% and 40$\sim$50\% \cite{Particledata}, respectively.
The $\pi \pi N$ decay occurs through three different modes,
i.e., $\pi \Delta$, $\rho N$ and $N(\pi \pi)_{\rm {s-wave}}^{I=0}$.
The $\pi \Delta$ channel is in either s-wave or d-wave state.
The branching fractions of the s-wave and d-wave decays are
5$\sim$12\% and 10$\sim$14\%, respectively.
The branching fraction to $\rho N$ is 15$\sim$25\%
and $N(\pi \pi)_{\rm {s-wave}}^{I=0}$ is almost negligible.
For simplicity, hereafter we denote $N^{*}(1520)$ as $N^{*}$.

Bhalerao {\it et al}.\cite{Liu} and Arima {\it et al}.\cite{Arima}
constructed the isobar model so as to describe the $\pi N$ scattering
in the D$_{13}$ channel.
In this model, the $N^{*}$ resonance is treated as the isobar state.
The self-energy of $N^{*}$ should include the contributions of both
$\pi N$ and $\pi \pi N$ channels as known from
the Particle Data\cite{Particledata}.
In their models, the $\pi \pi N$ channel was effectively regarded
as the $\pi \Delta$ channel.
We extend their models to apply to the pion photoproduction.
We explicitly include three important decay modes, i.e.,
s- and d-wave $\pi \Delta $ and $\rho N$, in the $\pi \pi N$ decay channel,
since each of such modes plays a significant role in the double pion
photoproduction process.
This will be discussed later in detail.

In the isobar model, the $\pi N$ $t$-matrix in the D$_{13}$ channel
is written as
\ben
\label{eq:D13}
t=
  \frac{ F_{ \pi N N^{*} } F_{ \pi N N^{*} }^{\dag} }
       {\sqrt{s} - M_{ N^{*} }^{0} -\Sigma _{ {\rm total} } },
\een
where $\sqrt{s}$ and $M_{ N^{*} }^{0}$ denote the total energy
in the center-of-mass system
and the bare mass of $N^{*}$, respectively. 
The total self-energy of $N^{*}$, i.e., $\Sigma _{ {\rm total} }$,
is expressed as 
\ben
\label{self:total}
    \Sigma_{ { \rm total } }
                =
     \Sigma _{ \pi N  }
     + \Sigma _{ \pi \Delta }^{s}
     + \Sigma _{ \pi \Delta }^{d}
     + \Sigma _{ \rho N  },
\een
where $\Sigma _{ \pi N  }$, $\Sigma _{ \pi \Delta }^{s}$, 
$\Sigma _{ \pi \Delta }^{d}$ and $\Sigma _{ \rho N  }$ are 
due to the coupling to the $\pi N $, s-wave $\pi \Delta$,
d-wave $\pi \Delta$ and $\rho N $ channels, respectively.

The vertex function for the $\pi N \rightarrow N^{*}$ is written as 
\ben
\label{op:piNNs}
F_{ \pi N N^{*} }^{\dag}
                       =
&&
                   -i 
                   (2\pi )^{3/2}
                     \sqrt{ 
                      \frac{ 2\omega_{\pi}(p) E_{N}(p)}
                           {            M             }
                           }
                     \frac
                         { f_{ \pi N N^{*} } }
                         { \sqrt{ 2(m+M) } }
                     \left(
                           \frac{ p }
                                { p_{\pi N N^{*}} }
                     \right)^{2}
                     e^{-(p / p_{\pi N N^{*}} )^{2} }
                  \left(
                        S^{(2)\dag }
                            \cdot
                        Y_{2}(\hat {\bm p } )
                  \right),
\een
where ${\bm p}$ and $\hat {\bm p}$ are the pion momentum
and its unit-vector in the $\pi N$
center-of-mass system,
respectively
and $\omega_{\pi}(p)=\sqrt{ m^{2}+|{\bm p}|^{2} }$,
$E_{N}(p)=\sqrt{ M^{2}+|{\bm p}|^{2} }$
and $p = |{\bm p}|$.
$f_{ \pi N N^{*} }$ is the $\pi N N^{*}$ coupling constant
and $p_{\pi N N^{*}}$ is the $\pi N N^{*}$ range parameter, 
and $M$ and $m$ denote nucleon and pion masses, respectively.
$S^{(2)\dag}$ in Eq.(\ref{op:piNNs}) is defined by 
\ben
S^{(2) \dag}
            =
             \sqrt{ \frac{ 2 }{ 5 } }
              \left[
                    {\bm S}^{\dag} \times {\bm \sigma} 
              \right]
                     ^{(2)},
\een
where ${\bm S}^{\dag}$ is the spin transition operator from 1/2 to 3/2 
and ${\bm \sigma}$ is the ordinary Pauli spin matrix.

The $N^{*}$ self-energy due to the coupling to the $\pi N$ channel is written as
\ben
\label{self:piNNs}
\Sigma_{\pi N}(\sqrt{s})
                        &=&
                \frac{ f_{\pi N N^{*}} ^{2} }
                                     {  2(m+M)  }
                      \int_{0}^{\infty}
                             {\rm d}p 
                      \frac{ p^{2}e^{ -2(p/p_{\pi N N^{*}})^{2}}  }
                            { \sqrt{s}-\omega_{\pi}(p)-E_{N}(p) +i\epsilon }
                               \left(
                                    \frac{ p }{ p_{\pi N N^{*}} }
                               \right) ^{4}.
\een
This expression is derived by using the vertex function of Eq.(\ref{op:piNNs}).

The $N^{*}$ self-energy due to the coupling to
the s-wave or d-wave $\pi \Delta$ channel
is expressed in a similar fashion, 
\ben
\label{self:piDsw}
\Sigma_{\pi \Delta}^{s(d)}(\sqrt{s})
        &=&
                \int
                \frac{ {\rm d}^{3}p }
                     { (2 \pi )^{3} }
     \frac{ 1 }{ 2\omega_{\pi}(p) }
     \frac{ 
            F_{ \pi \Delta N^{*} }^{s(d)}
            F_{ \pi \Delta N^{*} }^{s(d)\dag}
           }
           { 
       \sqrt{s}-\omega_{\pi}(p)
       -E_{\Delta}(p)
       -\Sigma_{\Delta}^{(\pi N)}(p,\sqrt{s})
           },
\een
where $E_{\Delta}=\sqrt{(M_{\Delta}^{0})^{2}+|{\bm p}|^{2}}$
and $M_{\Delta}^{0}$
is the bare mass of $\Delta$.
$\Sigma_{\Delta}^{(\pi N)}$ is the $\Delta$ self-energy due to the coupling
to the $\pi N$ channel,
which expression is given in Ref. \cite{Arima}.
We employ the same $\pi N \Delta$ vertex function and bare mass of $\Delta$
used by Betz and Lee\cite{BetzLee}.
The vertex functions for the $N^{*} \rightarrow \pi \Delta$ are defined as
\ben
\label{op:piDNS}
   F^{s\dag }_{\pi \Delta N^{*}}(p)
                    &=&
    -i
          (2\pi)^{3/2}
          \sqrt{
          \frac{2 \omega_{\pi}(p)}{ 2(m+M) }
               }
                f_{\pi \Delta N^{*}}^{s} 
     e^{-\left( p/p_{\pi \Delta N^{*}}^{s}\right)^{2}}
     Y_{00}(\hat{{\bm p}}),
\\
\label{op:piDND}
   F^{d\dag }_{\pi \Delta N^{*}}(p)
                &=&
                  -i(2\pi)^{3/2}
                  \sqrt{ \frac{2 \omega_{\pi}(p)}{  2(m+M) } }
                         f_{\pi \Delta N^{*}}^{d}
                  \left(
                  \frac{ p }{ p_{\pi \Delta N^{*}}^{d} }
                  \right)^{2}
                    e^{-(p/p_{\pi \Delta N^{*}}^{d})^{2} }
     \left(
     S_{3/2}^{(2)\dag}
     \cdot
     Y_{2}(\hat {\bm p})
     \right),
\een
where $s$ and $d$ denote s-wave and d-wave $\pi \Delta$ states,
and
$f_{\pi N^{*}\Delta}^{s}$ and $f_{\pi \Delta N^{*}}^{d}$ are
the s-wave and d-wave $N^{*} \rightarrow \pi \Delta$ 
coupling constants, respectively.
$p_{\pi \Delta N^{*}}^{s,d}$ are the $\pi \Delta N^{*}$ range parameters.
The spin transition operator from 3/2 to 3/2, $S_{3/2}^{(2)\dag}$,
in Eq.(\ref{op:piDND}) is defined by
\ben
  \left<
         \frac{3}{2} m' \left| S_{3/2 \mu} ^{(2)\dag}
          \right| \frac{3}{2} m
   \right>
          =
  \left(
         \frac{3}{2} m 2 \mu
        \left|
         \frac{3}{2} m'
         \right.
   \right),
\een
where $(j_{1}m_{1}j_{2}m_{2}|JM)$ is the corresponding Clebsch-Gordan
coefficient.

The $N^{*}$ self-energy due to the coupling to the
$\rho N$ channel is written as
\ben
\label{self:rhoNNs}
\Sigma_{\rho N}(\sqrt{s})
        &=&
               \int
               \frac{ {\rm d}^{3}p }
                    {(2 \pi )^{3} }
     \frac{ M }{ 2\omega_{\rho}(p)E_{N}(p) }
     \frac{ 
           F_{ \rho N N^{*} }
           F^{\dag}_{ \rho N N^{*} }
           }
          { 
            \sqrt{s}-\omega_{\rho}(p)
            -E_{N}(p)
            -\Sigma_{\rho \pi \pi}(p,\sqrt{s})
          },
\een
where $\omega_{\rho}(p)=\sqrt{ (m_{\rho}^{0})^{2}+|{\bm p}|^{2} }$
and $m_{\rho}^{0}$ is the bare mass of the $\rho $-meson . 
The vertex function for the $N^{*} \rightarrow \rho N$ is
\ben
\label{op:rhoNNs}
   F^{\dag}_{ \rho N N^{*} }
&=&
    (2 \pi )^{3/2}
          \sqrt{ \frac{ 2 \omega _{\rho}(p) E_{N}(p)}{ M } }
  f_{ \rho N N^{*} } e^{ -(p/p_{\rho N N^{*}})^{2} }
    \left(
    {\bm S^{\dag}}
      \cdot
    {\bm \varepsilon _{\rho} }
    \right)
     Y_{00}({\hat{\bm p}}),
\een
where ${\bm \varepsilon _{\rho}}$ is the $\rho$-meson polarization
vector and $p_{\rho N N^{*}}$ is the $\rho N N^{*}$ range parameter,
$f_{\rho N N^{*}}$ is the $\rho N N^{*}$ coupling constant.
$\Sigma_{\rho \pi \pi}$ in Eq.(\ref{self:rhoNNs}) is the
$\rho$-meson self-energy which is due to the coupling to the
$\pi \pi$ state.
The $\rho \pi \pi$ vertex function is assumed to take form:
\ben
\label{op:rhopipi}
F_{\rho \pi \pi}
                &=&
                  2h_{\rho}(\kappa)
                  ({\bm \varepsilon}_{\rho}\cdot {\bm \kappa}),
\\
h_{\rho}(\kappa) &=& \frac{ f_{\rho \pi \pi} }
                        {1 + (\kappa /q_{\rho \pi \pi })^{2} },
\een
where $q_{\rho \pi \pi}$ is the $\rho \pi \pi$ range parameter
and
$f_{\rho \pi \pi}$ is the $\rho \pi \pi$ coupling constant.
Using this vertex function, $\Sigma _{\rho \pi \pi }$ is written as
\ben
\label{eq:rho-pi-pi-self}
\Sigma_{\rho \pi \pi}(p,\sqrt{s})
                          & = &
                \frac{   1   }
                     {12\pi^{2} \omega_{\rho}(p) }
                \int_{0}^{\infty}
                      {\rm d}\kappa
                \frac{ \kappa ^{4} }{ \omega_{\pi}^{2}(\kappa) }
                \frac{ (h_{\rho}(\kappa) )^{2} }
                 { \sqrt{s}-E_{N}(p)-
                  \sqrt{ 4 \omega_{\pi}^{2}(\kappa) + p^{2} }
                     +i\epsilon }
		f_{\kappa},
\een
where
$f_{\kappa}=2 \omega_{\pi}(\kappa)/
\sqrt{ 4 \omega_{\pi}^{2}(\kappa) + p^{2} }$.
The above expression for the $\rho$-meson self-energy is obtained by extending
the self-energy in the rest frame of $\rho$ to that in the moving frame.
It should be noted that the imaginary part of Eq.(\ref{eq:rho-pi-pi-self})
has a right form of the half width for the $\rho$-meson with the momentum $p$,
\ben
\label{eq:width}
\frac{\Gamma_{\rho \pi \pi}(p,\sqrt{s})}
      {2}
                     &=&
                        \frac{ 1 }
                             { 24 \pi \omega_{\rho}(p) }
                        \frac{ \kappa_{e}^{3} }
                             { \omega_{\pi}(\kappa_{e}) }
                  \left(
                         h_{\rho}(\kappa_{e})
                  \right)^{2},
\een
where $\kappa_{e}$ satisfies
$\sqrt{s}-E_{N}(p)-\sqrt{4\omega_{\pi}^{2}(\kappa_{e})+p^{2}}=0$.
When $m_{\rho}^{0}$ is replaced by the on-shell $\rho$-meson mass
and $q_{\rho \pi \pi}$ is taken to be infinite,
Eq.(\ref{eq:width}) is reduced to the width of Ref.\cite{Oset}.
In order to calculate the self-energy of the $\rho$-meson,
one needs to know three independent parameters,
i.e., $f_{\rho \pi \pi}$, $q_{\rho \pi \pi}$ and $m_{\rho}^{0}$.
These parameters can not be determined uniquely by using the
mass and width of the $\rho$-meson.
Therefore, we 
treat the parameter $q_{\rho \pi \pi}$  as a free parameter and
vary it to reproduce the $\gamma p \rightarrow \pi^{+} \pi^{0}n$ data.
If $q_{\rho \pi \pi}$ is fixed,
$m_{\rho}^{0}$ and $f_{\rho \pi \pi}$ are determined
by the following condition,
\ben
\label{eq:condition}
     m_{\rho}^{\rm exp}
    -\left(
           m_{\rho}^{0}+\Sigma_{\rho \pi \pi}(0,m_{\rho}^{\rm exp}+M)
     \right)
\approx i 
          \frac{ 154 }{ 2 }~~[{\rm MeV}],
\een
where $m_{\rho}^{\rm exp} \approx$ 770 MeV.

The self-energies and bare masses of $\Delta$ and $\rho$
in the $N^{*}$-propagator are obtained through the data such as the
$\pi N$ P$_{33}$ scattering, the resonance energies and their
widths in a phenomenological way as mentioned above.
So there are nine parameters which have to be determined:
the coupling constants ($f_{\pi N N^{*} }$, $f_{\pi \Delta N^{*}}^{s}$,
$f_{\pi \Delta N^{*} }^{d} $, $f_{\rho N N^{*} }$),
the range parameters ($p_{\pi N N^{*}}$, $p_{\pi \Delta N^{*}}^{s}$,
$p_{\pi \Delta N^{*}}^{d}$, $p_{\rho N N^{*}}$)
and the bare mass ($M_{N^{*}}^{0}$).
We determine the parameters by fitting them to branching ratios,
the $N^{*}$ resonance energy, its width
and the energy dependence of the $\pi N$ D$_{13}$ scattering amplitude.
In our model, we use 1520 MeV as the resonance energy and
120 MeV as the width.
We take a fraction of 58\% for the decay into $\pi N$, 10\% into
s-wave $\pi \Delta$, 10\% into d-wave $\pi \Delta$ and
22\% decay into the $\rho N$ channel, respectively.

The parameters obtained are given in Table \ref{table:coupling}.
The parameter-set (I), (II) and (III) have been obtained by using
Eq.(\ref{op:rhopipi}) with $q_{\rho \pi \pi } = \infty$, 100 MeV/c
and 200 MeV/c, respectively.
In the parameter-set (I),
the self-energy and the bare mass of $\rho$ in Eq.(\ref{self:rhoNNs})
have been assumed to be the width and the on-shell mass, respectively,
which corresponds to the treatment of Ref.\cite{Oset}.
One finds that the $\pi N$ D$_{13}$
partial-wave amplitude can be equivalently reproduced by any parameter-set in
Table \ref{table:coupling}.
It should be noted that the sign of coupling constants
is not determined from the data,
since coupling constants appear as their squared form in the self-energy.
Which parameter-set and which sign are appropriate will be discussed
in Sec.\ref{sec:discussion}.
\section{$\gamma N N^{*}$ coupling}
\label{sec:photoncoupling}
In this section, we show how to determine the $\gamma N N^{*}$
coupling constants.
Obviously, the $\gamma N N^{*}$ vertex has two independent helicity 
couplings. 
For the proton target,
the helicity 1/2 amplitude is small enough compared with
the helicity 3/2 amplitude.
Hence, helicity 1/2 amplitude could be neglected\cite{Moorhouse,Fujii,Arndt}.
For the neutron target, on the other hand,
one can not use this approximation,
since the helicity 1/2 amplitude is non-negligible.

The resonant amplitude in the isobar model has the form
\ben
\label{eq:isobar}
   T_{ N^{*} }^{ \gamma N } =
                            F_{ \pi N N^{*} }
                            \frac{1}
                                 {\sqrt{s} - M_{ N^{*} } 
                                   -\Sigma _{ {\rm total} }}
                            F_{ \gamma N N^{*} }^{\dag},
\een
where $F_{ \gamma N N^{*} }$ is the vertex function
for the $\gamma N \rightarrow N^{*}$ transition. 
For the helicity 1/2 transition, $F_{\gamma N N^{*}}^{\dag}$
is written as
\ben
\label{op:h12}
 F^{1/2 \dag}_{\gamma N N^{*}}
                      &=& -i
           g_{1/2}
           \left(
              {\bm S}^{\dag} \cdot \hat {\bm k}
           \right)
              ({\bm \sigma} \cdot \hat {\bm k} \times
              {\bm \varepsilon}),
\een
where $g_{1/2}$ and ${\bm \varepsilon}$ are the helicity 1/2 coupling constant
and photon polarization vector, respectively, and $\hat{\bm k}$ denotes 
the unit vector of initial photon momentum.
In our approximation,
this helicity coupling constant for the proton is set to zero.
The helicity 3/2 transition $F_{\gamma N N^{*} }^{\dag}$ operator is written as
\ben
\label{op:h32}
   F^{3/2 \dag}_{\gamma N N^{*}}&=&
                   g_{3/2}
   \left\{
          \left( {\bm S}^{\dag}\cdot {\bm \varepsilon} \right)
         +\frac{i}{2}
          \left( {\bm S}^{\dag}\cdot \hat {\bm k} \right)
               ({\bm \sigma} \cdot 
		\hat {\bm k} \times {\bm \varepsilon}) 
\right \},
\een
where $g_{3/2}$ is the helicity 3/2 coupling constant.

The relevant multipole amplitudes can not be described by only the resonant
form of Eq.(\ref{eq:isobar}) since there is a non-negligible background process.
Actually, the full D$_{13}$ amplitude should be expressed as the sum of
two terms, i.e.,
the background and the $N^{*}$ resonant
terms (Fig.\ref{fig:D13FULL})\cite{KMO,Lee},
\ben
T(D_{13})&=&T_{B}+\tilde{T}_{ N^{*} },
\\
\label{eq:Ttilde}
    \tilde{T}_{ N^{ * } } & = &
                           F_{ \pi N N^{*} }
                            \frac{1}
                                 {\sqrt{s} - M_{ N^{*} } 
                                   -\Sigma _{ {\rm total} }}
                           \tilde{ F }_{ \gamma N N^{*} }^{1/2,(3/2)\dag},
\een
where $T_{B}$ is the non-resonant multipole amplitude which is
obtained from the partial-wave decomposition of the Born term in the
$\gamma N \rightarrow \pi N$ reaction.
The $\gamma N N^{*}$ vertex function is rewritten as
\ben
\label{eq:ftilde2}
    \tilde{ F }_{ \gamma N N^{*} }^{1/2 (3/2)\dag} 
               &=&
             \tilde{g}_{1/2 (3/2)}(\sqrt{s})
                        \frac{ F_{\gamma N N^{*}}^{1/2 (3/2)\dag} }
                             { g_{1/2 (3/2)} },
\een            
where $\tilde{g}_{1/2 (3/2)}$
represents the effective helicity 1/2 (3/2) coupling constant,
which includes the contribution of the $N^{*}$ production through
the $\pi N$, $\pi \Delta$ and $\rho N$ intermediate states
(Fig.\ref{fig:effectivecoupling} (c) to (e)) in addition to the direct
production (Fig. \ref{fig:effectivecoupling} (b)).
This coupling constant is complex and energy dependent.

The effective $\gamma N N^{*}$ coupling
constant $\tilde{g}_{1/2 (3/2)}$ is phenomenologically
determined by through a fit to the experimental multipole amplitudes of
Refs.\cite{Moorhouse,Fujii,Arndt}, instead of calculating the diagrams of
Fig.\ref{fig:effectivecoupling} (c),(d) and (e) in a microscopic way.
Here we use the Born term with the same coupling constants and cutoff
employed by Nozawa {\it et al}.\cite{Lee}.
If the multipole amplitude of Ref.\cite{Arndt} is used, for example,
the helicity 3/2 coupling constant for the proton target becomes
$\tilde{g}_{3/2}=0.1621+i0.0522$ at 750 MeV photon energy.
\section{Model of the $\gamma N\longrightarrow \pi \pi N$ reaction}
\label{sec:2p}
The total cross section for the 
$\gamma N \rightarrow \pi^{1} \pi^{2} N$ reaction is given by
\ben
\sigma
      &=&
             \frac{ 1 }{ 2 k }
             \frac{ M }{ E_{i} }
             \frac{ 1 }{ v_{s} }
         \int
             \frac{{ \rm d}^{3}p_{f}}{(2 \pi )^{3} }
             \frac{{ \rm d}^{3}q_{1}}{(2 \pi )^{3} }
             \frac{{ \rm d}^{3}q_{2}}{(2 \pi )^{3} }
             \frac{ M }{ E_{f} }
             \frac{ 1 }{ 2 \omega_{\pi}(q_{1}) }
             \frac{ 1 }{ 2 \omega_{\pi}(q_{2}) }
\nonumber
\\
\label{eq:crosssection}
     & &     
             \times     
             (2 \pi )^{4}
             \delta ^{(4)}
             (p_{i}+k-p_{f}-q_{1}-q_{2})
             \displaystyle\Sigma_{\nu \nu '}
             \frac{ 1 }{ 2 }
                          \left|
                          \left< 1/2, \nu
                          \left| T \right|
                                1/2, \nu ' \right>
                          \right| ^{2},
\een
where 
$p_{i}=(E_{i},{\bm p}_{i})$,
$p_{f}=(E_{f},{\bm p}_{f})$
and $q_{a}=(\omega _{\pi},{\bm q}_{a})$$(a=1,2)$ are the initial
nucleon, the final nucleon  and the final pion ($\pi^{1,2}$) 4-momenta 
in the center-of-mass system, respectively,
and $v_{s}$ is the relative velocity of the initial nucleon and the photon. 
The absolute square of the invariant matrix element $T$ for the
$\gamma N \rightarrow \pi \pi N$ reaction is summed over the
final nucleon spin states ($\nu '$) and averaged over the initial nucleon
spin states ($\nu$).

We describe how the matrix element $T$ in Eq.(\ref{eq:crosssection})
is derived within our approach.
We assume that the $\gamma N \rightarrow \pi \pi N$ reaction 
is dominated by the processes of the
$\gamma N \rightarrow \pi \Delta \rightarrow \pi \pi N$ and 
the $\gamma N \rightarrow N^{*} \rightarrow \pi \pi N$. 
In this assumption,
there are four important processes shown in
Figs.\ref{fig:2pi-1}(a),(b),(c) and (d).
We neglect other possible diagrams involving $\Delta$,
which are obtained from the requirement of the gauge invariance,
since these contributions has been shown to be small\cite{Oset}.

The diagram (a) in Fig.\ref{fig:2pi-1} contains 
the $\gamma N \pi \Delta $ contact term, i.e.,
the $\Delta$ Kroll-Ruderman term.
This $\gamma N \pi \Delta$ contact term operator $F_{\Delta \rm KR}^{\dag}$
is obtained from the strong $\pi N \Delta$ vertex function by requiring
the gauge invariance.
Instead of using the effective Lagrangian\cite{Oset},
we start from the vertex function with a form factor.
The $N \rightarrow \pi \Delta$ transition operator $F_{\pi N \Delta}^{\dag}$
is assumed to be the same form with the $\Delta \rightarrow \pi N$
vertex function which is phenomenologically given in Ref.\cite{BetzLee}.
Since the range parameter $Q_{\Delta}$ (see Appendix)
may not be necessarily the same,
we treat it as a free parameter and vary it to fit the
$\gamma p \rightarrow \pi^{+} \pi^{-} p$ cross section
(see Table \ref{table:coupling}).
Its matrix element in coordinate space in non-relativistic limit
is given by (suppressing the isospin factor)
\ben
\label{eq:piNm}
     \left<\pi \Delta
          \left|F_{\pi \Delta N}^{\dag}\right|
     N\right>
               &=&
                \int d^{3}r_{N}d^{3}\rho
                \left\{
                       \Psi_{\Delta}^{\dag}
                        ({\bm r}_{N}-\frac{m_{\pi}}{M_{\pi \Delta}}{\bm \rho})
                       \Phi_{\pi}^{\dag}
                        ({\bm r}_{N}+\frac{m_{\pi}}{M_{\pi \Delta}}{\bm \rho})
                \right\}
\nonumber
\\
                &&(+i)({\bm S}^{\dag}\cdot
                       {\stackrel{\leftarrow}{\nabla}_{\rho}})
                   H(\rho)
                  \Psi_{N}({\bm r}_{N}),
\een
where ${\bm r}_{N}$ and ${\bm \rho}$ are the $\pi \Delta$ center-of-mass
and relative coordinates, respectively,
and $M_{\pi \Delta}=M_{\Delta}+m$.
The $\pi N \Delta$ form factor $H$ is given in Appendix.
The $\Delta$ and pion wave functions in Eq.(\ref{eq:piNm})
may be expanded around ${\bm r}_{N}$
in power series of the relative coordinate ${\bm \rho}$.
If all gradients with respect to ${\bm r}_{N}$ operating on
the $\Delta$ and pion wave functions are replaced by
$\nabla _{N} -ie_{\Delta} {\bm A}({\bm r}_{N})$
or
$\nabla _{N} -ie_{\pi} {\bm A}({\bm r}_{N})$,
where
$e_{\Delta}$ and $e_{\pi}$ are the $\Delta$ and pion electric charges,
the resulting vertex function is invariant under
the gauge transformation\cite{Kumano}.
The electromagnetic interaction for the
$\gamma N \rightarrow \pi \Delta$ process is then derived by the expansion
to order ${\bm A}$.
Simplifying this interaction further according to the prescription
described in Ref.\cite{Kumano}, we obtain
\ben
\label{op:KR}
F_{\Delta \rm{KR}}^{\dag} &=& 
            -i
             \left\{
                    G_{1}
            ({\bm S}^{\dag}\cdot {\bm \varepsilon})
                        +
                    G_{2}
            ({\bm S}^{\dag}\cdot {\bm q}_{a})
            ({\bm q}_{a}\cdot {\bm \varepsilon})
             \right\}.
\een
This expression, i.e., the minimal interaction current,
is employed in our model.
Here, masses of the pion and the $\Delta$ in Eq.(\ref{eq:piNm}) are
replaced by their energies as the relativistic generalization
and the $\gamma N \pi \Delta$ form factors $G_{1}$, $G_{2}$ are
defined in Appendix.
The first term in the left-hand side of Eq.(\ref{op:KR})
corresponds to the ordinary contact term (the $\Delta$ Kroll-Ruderman term).
The second term, on the other hand,
appears due to the presence of the $\pi N \Delta$ form factor.
It should be noted that the above minimal interaction current has a
contribution to
the $\gamma p \pi^{0} \Delta ^{+}$ vertex because of the charged $\Delta$,
although there is no contribution within the framework of the effective
Lagrangian.
As shown later, this interaction has a small but non-negligible effect
on the $\gamma p \rightarrow \pi^{0} \pi^{0} p$ cross section.

In our model, the $\gamma N \pi \Delta$ pion-pole term $F_{\Delta \rm PP}^{\dag}$
included in Fig.\ref{fig:2pi-1}(b) will be derived from the time-ordered
perturbation theory.
The vertex functions for the
$\Delta \rightarrow \pi N$
and
$N \rightarrow \pi \Delta$
transitions are of the same form as that in Eq.(\ref{eq:piNm}).
But the range parameter $Q_{\Delta}$
of the latter is taken to be the same as that of Eq.(\ref{op:KR}).
The operator $F_{\Delta \rm PP}^{\dag}$ is then written as
\ben
\label{op:PP}
F_{\Delta \rm{PP}}^{\dag}&=&
          \frac{ig_{p}}{2\omega _{\pi}(|{\bm q}_{a}-{\bm k}|)}
         \left\{
          \frac{ H(\kappa _{1}) }
               { D_{\Delta}(q_{a},E_{N}(k)
                             -\omega _{\pi}(|{\bm q}_{a}-{\bm k}|)) }
             ({\bm S}^{\dag}\cdot {\bm \kappa _{1}})
         \right.
\nonumber
\\
         &&~~~~~~~~~~~~~~~~~~\left.
          -\frac{ H(\kappa _{2}) }
                { k-\omega _{\pi}(q_{a})-\omega _{\pi}(|{\bm q}_{a}-{\bm k}|) }
            ({\bm S}^{\dag}\cdot {\bm \kappa _{2}})
             \right\}
          ({\bm q}_{a}\cdot {\bm \varepsilon}),
\een
where
\ben
{\bm \kappa}_{1}&=&
                {\bm q}_{a}
                -\frac{ E_{\Delta}(q_{q}) {\bm k} }
                { E_{\Delta}(q_{a})+\omega _{\pi}(|{\bm q}_{a}-{\bm k}|) },
\\
{\bm \kappa}_{2}&=&
                      {\bm k}
                     -\frac{ E_{N}(k){\bm q}_{a} }
                           { E_{N}(k)+\omega _{\pi}(|{\bm q}_{a}-{\bm k}|) },
\\
D_{\Delta}(p,E)&=& 
E-E_{\Delta}(p)-\Sigma _{\Delta}^{(\pi N)}(p,E).
\een
Here, ${\bm \kappa}_{1}$ and ${\bm \kappa}_{2}$ are the
$\pi \Delta$ and $\pi N$ relative momenta in the intermediate state and
$\Sigma_{\Delta}^{(\pi N)}(p,E)$ is the self-energy of $\Delta$ with
the momentum $p$.
The charge-dependent factor $g_{p}$ is given in Appendix.

Using Eqs.(\ref{op:KR}) and (\ref{op:PP}),
we can write the invariant matrix element for the diagrams of 
Fig.\ref{fig:2pi-1}(a) and (b) in the following.
\ben
\label{amp:a}
T_{{\Delta \rm{KR}},({\rm{PP}})}
&=&
                \frac{ F_{\pi N \Delta} F_{\Delta \rm{KR,(PP)}}^{\dag} }
{
           \sqrt{s}
          -\omega_{\pi}(q_{a})
          -E_{\Delta}(q_{a})
          -\Sigma_{\Delta}^{(\pi N)}(q_{a},\sqrt{s})
},
\een
where $T_{\Delta \rm KR}$ and $T_{\Delta \rm PP}$ represent
the $\Delta$ Kroll-Ruderman term and $\Delta$ pion-pole term, respectively.
Multiplying Eq.(\ref{amp:a}) by appropriate isospin factors,
we get the $T$-matrix elements for various reactions.

For the $\gamma N \rightarrow N^{*} \rightarrow \pi \pi N$ process,
there are two possible processes accompanied with either
the $N^{*} \rightarrow \pi \Delta$ or $N^{*} \rightarrow \rho N$ decay
as shown in Figs.\ref{fig:2pi-1}(c) and (d).
The $N^{*}$ resonance can decay into both s-wave and d-wave
$\pi \Delta$ states.
In order to construct the $T$-matrix involving the $N^{*}$,
we use the strong and electromagnetic vertex functions obtained in
Secs.\ref{sec:Nstar} and \ref{sec:photoncoupling}.
Using Eqs.(\ref{op:piDNS}),
(\ref{op:piDND}), (\ref{op:rhoNNs}), (\ref{op:rhopipi})
and (\ref{eq:ftilde2}),
the $T$-matrix elements of Fig. \ref{fig:2pi-1}(c) and (d) are written as
\ben
\label{amp:c}
T_{N^{*}(\pi \Delta)}^{\rm{s(d)-wave}}
&=&
\frac{ F_{\pi N \Delta}
       F_{\pi \Delta N^{*}}^{s(d)\dag}
       {\tilde {F}}_{\gamma N N^{*}}^{\dag} }
     { 
(
               \sqrt{s}-\omega_{\pi}(q_{a})
      -E_{\Delta}(q_{a})-\Sigma_{\Delta}^{(\pi N)}(q_{a},\sqrt{s})
)
       \left( \sqrt{s}-M_{ N^{*} } -\Sigma_{ \rm{total} }
      \right) 
},
\\
\label{amp:d}
T_{N^{*}(\rho N)}
&=&
\frac{F_{\rho \pi \pi}
      F_{\rho N N^{*}}
      {\tilde {F}}_{\gamma N N^{*}}^{\dag}}
     { 2\omega_{\rho}(q_{\rho})
     \left(\sqrt{s}-\omega_{\rho}(q_{\rho})-E_{N}(q_{\rho})
      -\Sigma_{\rho \pi \pi}(q_{\rho},\sqrt{s})
       \right) 
       \left(\sqrt{s}-M_{N^{*}}-\Sigma_{\rm{total}} \right) },
\een
respectively.
Here, $q_{\rho}= |{\bm q}_{1}+{\bm q}_{2}|$.

Therefore, the invariant matrix element $T$ in Eq.(\ref{eq:crosssection})
is expressed as follows:
\ben
\label{eq:T1}
T= 
T_{\Delta {\rm KR}}
+
T_{\Delta {\rm PP}}
+
T_{N^{*}(\pi \Delta)}^{{\rm s-wave}}
+
T_{N^{*}(\pi \Delta)}^{{\rm d-wave}}
+
T_{N^{*}(\rho N)}.
\een
\section{Numerical results and Discussion}
\label{sec:discussion}
In this section,
we present our calculations of the total cross sections for the
$\gamma N \rightarrow \pi \pi N$ reaction, which are shown in
Figs.\ref{fig:pmp-pmn} to \ref{fig:zzp-zzn}. 
We calculated them with the parameters obtained in previous sections,
but the sign of some strong coupling constants and
the range parameter $Q_{\Delta}(N \rightarrow \pi \Delta)$ will be
determined so as to
fit them to the $\gamma p \rightarrow \pi^{+} \pi^{-} p$ data.
Then, we will discuss which parameters of the
$\rho \pi \pi$ form factor are favored by the
$\gamma p \rightarrow \pi^{+} \pi^{0} n$ data.
In our numerical calculations,
the Monte Carlo integration package BASES25 \cite{MC} is used. 

At first, we show the results of 
$\gamma p \rightarrow \pi^{+} \pi^{-} p$
and $\gamma n \rightarrow \pi^{+} \pi^{-} n$ reaction cross sections
(solid lines) in Figs.\ref{fig:pmp-pmn}-(i) and (ii) which are calculated
with the parameter-set (I) in Table \ref{table:coupling}.
As can be seen from Fig.\ref{fig:pmp-pmn}-(i), 
the $\Delta$ Kroll-Ruderman term $T_{\Delta \rm{KR}}$
and $\Delta$ pion-pole term $T_{\Delta \rm{PP}}$ terms (dashed line)
dominate on the
$\gamma p \rightarrow \pi^{+} \pi^{-} p$ reaction.
We observe that the $N^{*}$ contribution (dash-dotted line)
alone is small
but the interference between the $N^{*}$ term
$T_{N^{*}(\pi \Delta)}^{\rm{s-wave}}$ and $T_{\Delta \rm{KR}}$ is important.
This strong interference occurs due to the fact that
$T_{N^{*}(\pi \Delta)}^{\rm{s-wave}}$ has the same structure as
$T_{\rm{KR}}$.
Because of this, the $N^{*}$ excitation is regarded as an important ingredient
in the $\gamma p \rightarrow \pi \pi N$ reaction. 

As mentioned in Sec.\ref{sec:Nstar},
there is an ambiguity about the sign of strong coupling constants,
$f_{\pi \Delta N^{*}}^{s,d}$ and $f_{\rho N N^{*}}$. 
We adopt a positive sign for $f_{\pi \Delta N^{*}}^{s}$ 
which gives rise to a constructive interference between
$T_{N^{*}(\pi \Delta)}^{\rm{s-wave}}$ and $T_{\Delta \rm{KR}}$.
The peak position of the $\gamma p \rightarrow \pi^{+} \pi^{-} p$ cross section
can be reproduced with this choice as shown in Fig.\ref{fig:pmp-pmn}-(i).
We find that the curve with
$f_{\pi \Delta N^{*}}^{s}>0$,
$f_{\pi \Delta N^{*}}^{d}<0$, and $f_{\rho N N^{*}}<0$
agrees well with the $\gamma p \rightarrow \pi^{+} \pi^{-} p$
data\cite{doublepi-exp1,doublepi-exp2,PIA,CAR}.
The range parameter $Q_{\Delta}( N \rightarrow \pi \Delta)$
is taken to be 420 MeV/c.
Hereafter, we will use the same sign for the coupling constants.
Furthermore,
the $\gamma n \rightarrow \pi^{+} \pi^{-} n$ reaction
cross section (Fig.\ref{fig:pmp-pmn}-(ii)) is calculated by
using the same parameter-set (I)
except for the $\gamma N N^{*}$ coupling.
Our calculation with this parameter-set is also in good agreement
with the $\gamma n \rightarrow \pi^{+} \pi^{-} n$ data\cite{PIA,CAR}.
This is different from the result by Tejedor {\it et al}.\cite{Oset}.
Their model could not reproduce the $\gamma n \rightarrow \pi^{+} \pi^{-} n$
data in spite of good agreement with the
$\gamma p \rightarrow \pi^{+} \pi^{-} p$ data.
This difference may be attributed mainly to the theoretical treatment of the
$\Delta$ Kroll-Ruderman term.
In any case, new experiments for the neutron target
would be welcome in order to check the validity of our model.

Secondly, we show the results of the
$\gamma p \rightarrow \pi^{+} \pi^{0} n$
and $\gamma n \rightarrow \pi^{-} \pi^{0} p$ reactions 
in Figs.\ref{fig:pzn-mzp}-(i) and (ii) (thin-solid lines),
which are calculated with the parameter-set (I).
In both reactions, we find a large discrepancy with the
data\cite{doublepi-exp2,PIA,CAR}.
Our model underestimates cross sections about a factor of two
compared with the data.
In these reactions, as shown in Fig.\ref{fig:pzn-mzp}-(i) and (ii),
$T_{\Delta \rm{KR}}$ and $T_{\Delta \rm{PP}}$ contribution
(short-dashed line) is very small.
We also find that the contribution of the $N^{*}$ terms
(dash-dotted line) is almost the same as that in the
$\gamma p \rightarrow \pi^{+} \pi^{-}p$ reaction
but there is no characteristic energy-dependence due to the
interference which is clearly observed in the
$\gamma p \rightarrow \pi^{+} \pi^{-} p$ cross section.
This can be understood by the isospin factor.
The isospin ratio of the $T_{\Delta \rm{KR}(PP)}$ term of the
$\gamma p \rightarrow \pi^{+} \pi^{0} n$ reaction to the 
$\gamma p \rightarrow \pi^{+} \pi^{-} p$ reaction is
\ben
\label{rel:KP}
\frac{
T_{\Delta \rm{KR}(PP)}(\pi^{+} \Delta^{0}\rightarrow \pi^{+} \pi^{0} n)
     }
      {
T_{\Delta \rm{KR}(PP)}(\pi^{-} \Delta^{++}\rightarrow \pi^{-} \pi^{+} p)
     }
       =
-\frac{\sqrt{2}}{3},
\een
where only the intermediate and final states are written in the invariant
matrix element $T$ but the initial state $\gamma N$ is omitted.
For instance, $T(\pi^{-} \Delta^{++} \rightarrow \pi^{+} \pi^{-}p)$
means
$T(\gamma p \rightarrow \pi^{-} \Delta^{++} \rightarrow \pi^{+} \pi^{-}p)$.
Eq.(\ref{rel:KP}) indicates that the cross section for the
$\gamma p \rightarrow \pi^{+} \pi^{0} n$ reaction is 2/9 times smaller than
$\gamma p \rightarrow \pi^{+} \pi^{-} p$ reaction.
For the $\Delta$ Kroll-Ruderman term, this relation holds exactly
if only the dominant term is considered, but
this feature is not changed even if  other small terms are included.
The isospin ratio of the $T_{N^{*}(\pi \Delta)}^{\rm{s(d)-wave}}$ term
of the $\gamma p \rightarrow \pi^{+} \pi^{0} n$ to 
$\gamma p \rightarrow \pi^{+} \pi^{-} p$ reaction is
\ben
\label{rel:NS}
\frac{
T_{N^{*}(\pi \Delta)}^{\rm{s(d)-wave}}
(\pi^{+}\Delta^{0}\rightarrow \pi^{+}\pi^{0}n)
     }
      {
T_{N^{*}(\pi \Delta)}^{\rm{s(d)-wave}}
(\pi^{-}\Delta^{++}\rightarrow \pi^{-}\pi^{+}p)
     }
       =
\frac{ 1 }{\sqrt{2}}.
\een
Eqs.(\ref{rel:KP}) and (\ref{rel:NS}) show that 
a relative sign between 
$T_{\rm{KR}(PP)}$ and $T_{N^{*}(\pi \Delta)}^{\rm{s-wave}}$
in the $\gamma p \rightarrow \pi^{+} \pi^{0} n$
reaction is different from that in the
$\gamma p \rightarrow \pi^{+} \pi^{-} p$ reaction,
and Eq.(\ref{rel:NS}) indicates that
$T_{N^{*}(\pi \Delta)}^{\rm{s-wave}}$ in the
$\gamma p \rightarrow \pi^{+} \pi^{0} n$ reaction is smaller than that in
$\gamma p \rightarrow \pi^{+} \pi^{-} p$ reaction.
Therefore, the interference between
$T_{\rm{KR}}$ and $T_{N^{*}(\pi \Delta)}^{\rm{s-wave}}$
in the $\gamma p \rightarrow \pi^{+} \pi^{0} n$ reaction is different
from that in the $\gamma p \rightarrow \pi^{+} \pi^{-} p$ reaction.
In addition,
there is another important feature regarding the $\rho$-production amplitude
(diagram (d) in Fig.\ref{fig:2pi-1}).
The isospin ratio of the $T_{N^{*}(\rho N)}$ term is
\ben
\label{rel:RHO}
\frac{
T_{N^{*}(\rho N)}
(\rho^{+}n \rightarrow \pi^{+}\pi^{0}n)
}
{
T_{N^{*}(\rho N)}
(\rho^{0}p \rightarrow \pi^{-}\pi^{+}p)
}
= \sqrt{2}.
\een
The Eq.(\ref{rel:RHO}) shows that the $T_{N^{*}(\rho N)}$ term 
in the $\gamma p\rightarrow \pi^{+}\pi^{0}n$ reaction
is larger than
that in the $\gamma p\rightarrow \pi^{+}\pi^{-}p$ reaction.
Hence, the $\rho$-meson production term is important
in the $\gamma p\rightarrow \pi^{+}\pi^{0}n$ reaction.
The large shift of the peak
compared with the calculation of the $\gamma p \rightarrow \pi^{+} \pi^{-}p$
cross section may be due to the large $T_{N^{*}(\rho N)}$ term.
The same arguments remain true for the
$\gamma n \rightarrow \pi^{+} \pi^{-} n$ and
$\gamma n \rightarrow \pi^{-} \pi^{0} p$ reactions.

As can be seen from
Figs.\ref{fig:pmp-pmn}-(i),(ii),
Figs.\ref{fig:pzn-mzp}-(i) and (ii)(thin-solid lines),
one finds that our model is successful
in the $\gamma N \rightarrow \pi^{+} \pi^{-} N$ reaction
but fails to reproduce the experimental results of the
$\gamma p \rightarrow \pi^{+} \pi^{0} n$ and
$\gamma n \rightarrow \pi^{-} \pi^{0} p$ reactions.
This difficulty concerning the $\pi^{0}$ production has been already
pointed out in other studies\cite{Oset,Laget}.
In order to improve our model and resolve this problem,
we examine the two possible processes shown in
Figs.\ref{fig:2pi-1} (e) and (f),
which may contribute more effectively to 
the $\gamma p \rightarrow \pi^{+} \pi^{0} n$ than 
the $\gamma p \rightarrow \pi^{+} \pi^{-} p$ reaction.
The diagram (e) in Fig.\ref{fig:2pi-1} is the final state rescattering process.
But we found from our rough estimate that the
cross section for this rescattering diagram (e) is very small.

Next, let us discuss the effect of the diagram (f) in Fig.\ref{fig:2pi-1}
in detail.
This diagram contributes only to the
$\gamma p \rightarrow \pi^{+} \pi^{0}n$
and $\gamma n \rightarrow \pi^{-} \pi^{0}p$ reactions,
since it contains $\gamma  N \rho N $ contact interaction
($\rho $-Kroll-Ruderman term).
The invariant matrix element for this diagram is written as
\ben
\label{amp:f}
T_{\rho \rm KR}
               &=&
\frac{ F_{\rho \pi \pi}F_{\gamma N \rho N}^{\dag} }
{ 2\omega_{\rho}(q_{\rho})(\sqrt{s}-\omega_{\rho}(q_{\rho})-E_{N}(q_{\rho})
          -\Sigma_{\rho \pi \pi}(q_{\rho},\sqrt{s}))
                        }.
\een
Here $F_{\gamma N \rho N}^{\dag}$ represents the $\gamma N \rho N$ contact term
(suppressing the isospin factor)
\ben
\label{op:rhocontact}
F_{\gamma N \rho N}^{\dag}
                  &=&
                      ief_{c}
                      \sqrt{ \frac{ E_{N}(k) + M}{ 2 M } }
                            \frac{ G^{T} }
                             { 2 M }
                        {\bm \varepsilon}_{\rho}
                         \cdot
                        ({\bm \sigma}\times {\bm \varepsilon}),
\een
where $G^{T}= -17.6$ is the tensor coupling to the $\rho N N$ channel
and the factor $f_{c}$ is taken to be
$\sqrt{m_{\rho}^{0}m_{\rho}'/(m_{\rho}^{\rm exp})^{2}}$.
$m_{\rho}'=m_{\rho}^{0} +\Sigma_{\rho \pi \pi}(0,M)$
is the mass of $\rho$ carrying zero momentum and zero energy.
The factor $f_{c}$ arises from the fact that the intermediate $\rho$-meson
mass in our model is different from the on-shell mass.
The effect of this factor is however small so that it does not affect
our final results significantly.
Thus, our new invariant matrix element $T$ matrix becomes
\ben
\label{eq:T2}
T= 
T_{\Delta {\rm KR}}
+
T_{\Delta {\rm PP}}
+
T_{N^{*}(\pi \Delta)}^{{\rm s-wave}}
+
T_{N^{*}(\pi \Delta)}^{{\rm d-wave}}
+
T_{N^{*}(\rho N)}
+
T_{\rho \rm KR}.
\een
We neglect the other possible diagram obtained by requiring the gauge
invariance, i.e., the $\rho$-meson pole diagram,
since its effect was found to be negligible.

In order to see the effect of the diagram (f) in Fig.\ref{fig:2pi-1},
we calculated the $\gamma p \rightarrow \pi^{+} \pi^{0} n$ cross section
by using the invariant matrix element $T$ of Eq.(\ref{eq:T2}), which is shown
as dotted line in Fig.\ref{fig:pzn-mzp}-(i).
In this calculation, we used the parameter-set (I).
As far as $T_{\rho \rm KR}$ term is concerned, in this case,
the theoretical treatment in our model is essentially the same with
the model by Murphy {\it et al}.\cite{Laget}.
From the comparison between the calculation with $T_{\rho \rm KR}$
(dotted line) and without $T_{\rho \rm KR}$ (thin-solid line),
the contribution of the diagram (f) in Fig.\ref{fig:2pi-1}
is found to be very small, which is consistent with the result of
Ref.\cite{Laget}.

To improve our model, we will treat the $\rho$-meson propagator in a
dynamical way since the $\rho$-meson involved in the double pion
photoproduction is not on-shell below the energy range 800 MeV. 
We allow the range parameter of the $\rho \pi \pi$ form factor to be
finite as described in Sec. \ref{sec:Nstar}.
We calculate the cross section with the
parameter-set (II) in Table \ref{table:coupling}
where the range parameter of the $\rho \pi \pi$ form factor
$q_{\rho \pi \pi}$ is 100 MeV/c
and 
$Q_{\Delta}(N \rightarrow \pi \Delta)$ is 400 MeV/c.
For the $\gamma p \rightarrow \pi^{+} \pi^{-} p$ cross section,
the result with the parameter-set (II) is almost the same as that with
the parameter-set (I).
As can be seen from Fig.\ref{fig:pzn-mzp}-(i), however,
the significant enhancement occurs in the cross section of the
$\gamma p \rightarrow \pi^{+} \pi^{0}n$ reaction (bold-solid line).
Our improved model
fairly well reproduces the data of the
$\gamma p \rightarrow \pi^{+} \pi^{0} n$ reaction,
except for the energy region above 750 MeV.
This result is quite different from those
by Tejedor {\it et al}.\cite{Oset} and Murphy {\it et al}.\cite{Laget}.
Furthermore, we have calculated
the $\gamma n \rightarrow \pi^{-} \pi^{0}p$ reaction cross section
by using the same parameter-set.
The calculation is shown in Fig.\ref{fig:pzn-mzp}-(ii)(bold-solid line).
We find that our model with this parameter-set is also able to reproduce
the experimental data for the neutron target\cite{PIA,CAR}.

In order to find the reason
why such significant enhancement has occurred,
we plot a 2$\pi$-spectral function, i.e.,
${\rm Im}\Sigma_{\rho \pi \pi}/|\sqrt{s}-m_{\rho}-\Sigma_{\rho \pi \pi}|^{2}$
which is proportional to the integrated cross section for the diagram (f).
The curves with the range parameter
$q_{\rho \pi \pi}=100$ (solid line), 200 (dashed line),
300 (dash-dotted line) MeV/c,
respectively, are plotted in Fig.\ref{fig:pipiamp}.
The peak is clearly seen at the $\rho$-meson resonance energy.
Furthermore, below the 400 MeV, 
there is a small bump in the curve with $q_{\rho \pi \pi} = 100$ MeV/c.
This non-negligible effect may cause a significant enhancement to the
$\gamma p \rightarrow \pi^{+} \pi^{0} n$ reaction cross section.
One can see the similar enhancement in the 2$\pi$-spectral function associated
with the $N{\bar N} \rightarrow 2\pi$ reaction in Ref.\cite{Toki}.
This low-energy behavior comes from the background contribution in the
isospin $I=1$ channel.
Our model for the 2$\pi$ scattering includes only the $\rho \pi \pi$
coupling, but not the background interaction.
However, the small range parameter in our model might simulate the
background interaction effectively.

We show the results of double neutral pion photoproduction processes
in Figs.\ref{fig:zzp-zzn}-(i) and (ii).
As can be seen from Fig.\ref{fig:zzp-zzn}-(i),
the peak position of the calculated cross section
of the $\gamma p \rightarrow \pi^{0} \pi^{0} p$
is in good agreement with the data\cite{doublepi-exp2}.
On the other hand, the magnitude of the cross section is underestimated
about a factor of two.
Murphy {\it et al}. suggested\cite{Laget} that
the $\gamma p \rightarrow \pi^{0} \pi^{0}p$
reaction cross section was enhanced by the presence of the P$_{11}$(1440)
resonance which decays into the $\sigma$-meson.
However, it seems that other important mechanisms are still missing.
We leave this problem as a next step since we are interested in
the total photoabsorption at present.
We also show the $\gamma n \rightarrow \pi^{0} \pi^{0} n$ reaction
cross section in Fig.\ref{fig:zzp-zzn}-(ii).
Unfortunately,
there are no experimental data to compare with our calculation.

Finally, we show the total photoabsorption cross sections (solid lines)
for $\gamma p$ and $\gamma n$ reactions
in Figs.\ref{fig:total-p-n}-(i) and (ii), respectively.
We assume that the total photoabsorption off a nucleon is dominated by
the $\gamma N \rightarrow \pi N$ and $\gamma N \rightarrow \pi \pi N$ reactions
in the energy region where we discuss.
The $\gamma N \rightarrow \pi N$ cross sections (dashed lines)
are calculated by using  the amplitude given in Ref.\cite{Arndt}
except for the D$_{13}$ amplitude.
The D$_{13}$ amplitude is treated in the same way described in 
Secs.\ref{sec:Nstar} and \ref{sec:2p}.
On the other hand,
the $\gamma N \rightarrow \pi \pi N$ reaction cross sections (dash-dotted lines)
are calculated by using our model with the parameter-set (II).
We found our model reproduces the experimental data
both $\gamma p$\cite{Nucl4,Arm-p} and $\gamma n$\cite{Arm-n} reactions
over the wide range of the energy.
We should mention that the double neutral pion photoproduction scarcely
contribute to the total photoabsorption cross section. 
\section{Conclusion}
\label{sec:conclusion}
We have constructed a simple model for the
$\gamma N \rightarrow \pi \pi N$ reaction.
It is assumed in this model that the processes of
$\gamma N \rightarrow \pi \Delta (1232)$,
$\gamma N \rightarrow N^{*}(1520)$
and $\gamma N \rightarrow \rho N$ are
dominant in the double pion production.
We treat the resonances such as $\Delta$, $N^{*}$ and $\rho$-meson
in a dynamical way.

The $\gamma N \pi \Delta$ contact operator
is derived from the strong $\pi N \Delta$ vertex function
by requiring the gauge invariance,
instead of the effective Lagrangian.
The range parameter of the $N \rightarrow \pi \Delta$ form factor
in this operator is determined so as to reproduce the
$\gamma p \rightarrow \pi^{+} \pi^{-} p$ data.
For the $\Delta$ resonance,
the dynamical model by Betz and Lee\cite{BetzLee} was used. 
We have carefully discussed the dynamical model for the $N^{*}$ resonance
because this resonance plays very important roles in the energy range
where we concern.
The strong vertex functions of
$\pi N N^{*}$, $\pi \Delta N^{*}$ and
$\rho N N^{*}$ are obtained from the $\pi N$ scattering amplitudes,
decay widths and the resonance energy of $N^{*}$.
The sign of each strong coupling constant is not determined from only the
above data.
However, as the sign affects the energy dependence of the double pion
photoproduction cross section, it could be fixed to reproduce the
$\gamma p \rightarrow \pi^{+} \pi^{-} p$ data.
The electromagnetic couplings of $\gamma N N^{*}$
are determined through a fit to the
$\gamma  N \rightarrow \pi N$ D$_{13}$ helicity 1/2 and 3/2
partial-wave amplitude.

Our model with the above parameters can simultaneously reproduce
total cross sections of both the 
$\gamma p \rightarrow \pi^{+} \pi^{-} p$ and
$\gamma n \rightarrow \pi^{+} \pi^{-} n$ reactions.
We found that the $\Delta$ Kroll-Ruderman term and the $\Delta$ pion-pole term
had a dominant contribution to these reactions and the interference
between the $\Delta$ Kroll-Ruderman term and the $N^{*}$ term
was very important to explain the energy-dependence of the cross sections.

For the $\gamma p \rightarrow \pi^{+} \pi^{0} n$ and
$\gamma n \rightarrow \pi^{-} \pi^{0} p$ reactions,
on the other hand,
the contributions of the $\Delta$ Kroll-Ruderman term and
the $\Delta$ pion-pole term are smaller compared with the 
$\gamma N \rightarrow \pi^{+} \pi^{-} N$ reaction.
In order to reproduce the $\gamma p \rightarrow \pi^{+} \pi^{0} n$ data,
we found that one should treat the $\rho$-meson in a dynamical model
where the $\rho \pi \pi$ vertex has a finite-ranged form factor.
In fact, it has turned out in our calculation with the appropriate range
parameter that the $\gamma N \rho N$ contact term contributes to these
$\pi^{0}$ productions significantly.
As a result, our improved model could simultaneously reproduce both
the $\gamma p \rightarrow \pi^{+} \pi^{0} n$
and $\gamma n \rightarrow \pi^{-} \pi^{0} p$
cross sections, except for the energy region above 750 MeV.
The disagreement at higher energy may be due to the fact that
other higher resonances are not taken into account in our model.

The total cross section of the $\gamma p \rightarrow \pi^{0} \pi^{0}p$
reaction is underestimated about a factor of two.
Still some important mechanisms are missing in our model.
We leave this problem as a next step.

Finally, we have calculated the total cross sections of the photoabsorption 
off a nucleon.
We found that our model is able to reproduce the experimental results of the
proton target as well as the neutron target.
The defect of our model regarding the double neutral pion production
gives little influence on the total photoabsorption cross section,
since the magnitude of its cross section is very small.
As far as the total cross section is concerned, we consider that our model
has a predictable power for nuclear processes.
Based on our model, we are investigating the mechanism which
has caused the resonance damping in the nuclear photoabsorption\cite{future}.
\acknowledgements
We are gratefull to Dr. N. Aizawa for instructing us BASES25.
We also thank Prof. Y. Sumi for indication of nuclear photoabsorption
data.
\appendix
\renewcommand{\thesection}{Appendix \Alph{section}}
\renewcommand{\thesection}{\Alph{section}}
\section{The $\pi N \Delta$ and $\gamma N \pi \Delta$ form factors}
\label{appendix:b}
\renewcommand{\thesection}{\Alph{section}}
The $\pi N \Delta $ form factor $H$ is defined as
\ben
H(q)=   \sqrt{ 6 \pi^{2} }
        \sqrt{\frac{ 2\omega_{\pi}(q) E_{N}(q) }{ M } }
        \frac{ g_{\pi N \Delta}(q)}{ q },
\een
where $q$ is the 3-momentum in the $\pi N$ center of mass system and
$g_{\pi N \Delta}$ is given by \cite{BetzLee}
\ben
\label{apeq:g}
g_{\pi N \Delta}=
                  \frac{ F_{\Delta} }
                       { \sqrt{2(m+M)} }
                  \frac{ q }
                       { m }
                  \left(
                  \frac{ Q_{\Delta}^{2} }
                       { Q_{\Delta}^{2}+q^{2} }
                  \right)^{2},
\een
where $F_{\Delta}$ is the coupling constant and $Q_{\Delta}$ is
the range parameter.
The $\gamma N \pi \Delta$ form factor $G_{1}$ and $G_{2}$
in Eq.(\ref{op:KR}) and the factor $g_{p}$ in Eq.(\ref{op:PP})
are given as follows.
\\
 1) $\gamma p \rightarrow \pi^{-} \Delta^{++}$ reaction
\ben
G_{1}&=&
        e\left\{Z_{2}H(|{\bm q}-Z_{2}{\bm k}|)
                +2Z_{3}H(|{\bm q}+Z_{3}{\bm k}|)\right\},
\\
G_{2}&=&
       e(-\tilde{h}_{\pi}+2\tilde{h}_{\Delta}),
\\
g_{p}&=& -2e.
\een
 2) $\gamma p \rightarrow \pi^{+} \Delta^{0}$ reaction
\ben
G_{1}&=&
        -\frac{e}{\sqrt{3}}Z_{2}H(|{\bm q}-Z_{2}{\bm k}|),
\\
G_{2}&=&
        -\frac{e}{\sqrt{3}}\tilde{h}_{\pi},
\\
g_{p}&=& \frac{ 2 }{\sqrt{3} }e.
\een
 3) $\gamma p \rightarrow \pi^{0} \Delta^{+}$ reaction
\ben
G_{1}&=&
       -\frac{e}{\sqrt{3}}Z_{3}H(|{\bm q}+Z_{3}{\bm k}|),
\\
G_{2}&=&
       -\sqrt{\frac{2}{3}}e\tilde{h}_{\Delta},
\\
g_{p} &=& 0
\een
 4) $\gamma n \rightarrow \pi^{+} \Delta^{-}$ reaction
\ben
G_{1}&=&
        e\left\{-Z_{2}H(|{\bm q}-Z_{2}{\bm k}|)
                -Z_{3}H(|{\bm q}+Z_{3}{\bm k}|)\right\},
\\
G_{2}&=&
        e(\tilde{h}_{\pi}-\tilde{h}_{\Delta}),
\\
g_{p}&=& 2e.
\een        
 5) $\gamma n \rightarrow \pi^{-} \Delta^{+}$ reaction
\ben
G_{1}&=&
       \frac{e}{\sqrt{3}}\left\{
        Z_{2}H(|{\bm q}-Z_{2}{\bm k}|)+Z_{3}H(|{\bm q}+Z_{3}{\bm k}|)
        \right\},
\\
G_{2}&=&
        \frac{e}{\sqrt{3}}(\tilde{h}_{\Delta}-\tilde{h}_{\pi}),
\\
g_{p}&=& -\frac{ 2 }{\sqrt{3} }e.
\een
where $e$ is electromagnetic charge and including isospin Clebsch-Gordan
coefficients of the $\gamma N \pi \Delta$ vertex and ${\bm q}$ is the
3-momentum of the out-going pion and
\ben
Z_{2}&=& \frac{E_{\Delta}(q)}{E_{\Delta}(q)+\omega (q)},
\\
Z_{3}&=& \frac{\omega(q)}{E_{\Delta}(q)+\omega(q)},
\\
\tilde{h}_{\pi}&=&
                     \frac{H(|{\bm q}-Z_{2}{\bm k}|)-H(q)}
                          {{\bm k}\cdot ({\bm {q}}-Z_{2}{\bm k}/2)},
\\
\tilde{h}_{\Delta}&=&
                     \frac{H(|{\bm q}+Z_{3}{\bm k}|)-H(q)}
                          {{\bm k}\cdot ({\bm {q}}+Z_{3}{\bm k}/2)}.
\een

\newpage
\mediumtext
\begin{table}[H]
\caption{
The parameters used in our model.
Only the absolute value of coupling constants is given.
The sign is discussed in the text.
}
\label{table:coupling}
\begin{center}
\begin{tabular}{|c||c||c||c|}
 & parameter-set (I) & parameter-set (II) & parameter-set (III)
\\ 
\hline
$ M_{N^{*}}${\footnotesize(MeV)} & 1597 & 1554 & 1566 
\\
\hline
$f_{\pi N N^{*}} $         &  1.09    & 1.13 & 1.13
\\ 
$p_{\pi N N^{*}}$ {\footnotesize(MeV/c)}       &  450    & 400 & 400
\\ 
\hline
$f_{\pi \Delta N^{*}}^{s}$ &  0.992   & 0.992 & 0.992
\\ 
$p_{\pi \Delta N^{*}}^{s} $ {\footnotesize(MeV/c)}      &  200    & 200 & 200
\\ 
\hline
$f_{\pi \Delta N^{*}}^{d}$ &  0.984   & 1.00 & 1.00
\\ 
$p_{\pi \Delta N^{*}}^{d}$ {\footnotesize(MeV/c)}       &  200    & 300 & 300
\\ 
\hline
$f_{\rho N  N^{*}}$                 &  1.56   & 0.928 & 0.583
\\ 
$p_{\rho N N^{*}}$ {\footnotesize(MeV/c)}       &  200    & 200   & 300
\\ 
\hline
$f_{\rho \pi  \pi}$      & 6.14 & 82.0 & 25.6
\\ 
$q_{\rho \pi \pi}$ {\footnotesize(MeV/c)}       & $\infty$  & 100   & 200
\\ 
\hline
{\footnotesize $Q_{\Delta}(N \rightarrow \pi \Delta)$ (MeV/c)}& 420 & 400 &400
\\ 
{\footnotesize $Q_{\Delta}(\Delta \rightarrow \pi N)$ (MeV/c)}& 358 & 358 &358
\end{tabular}
\end{center}
\end{table}
\newpage
\twocol
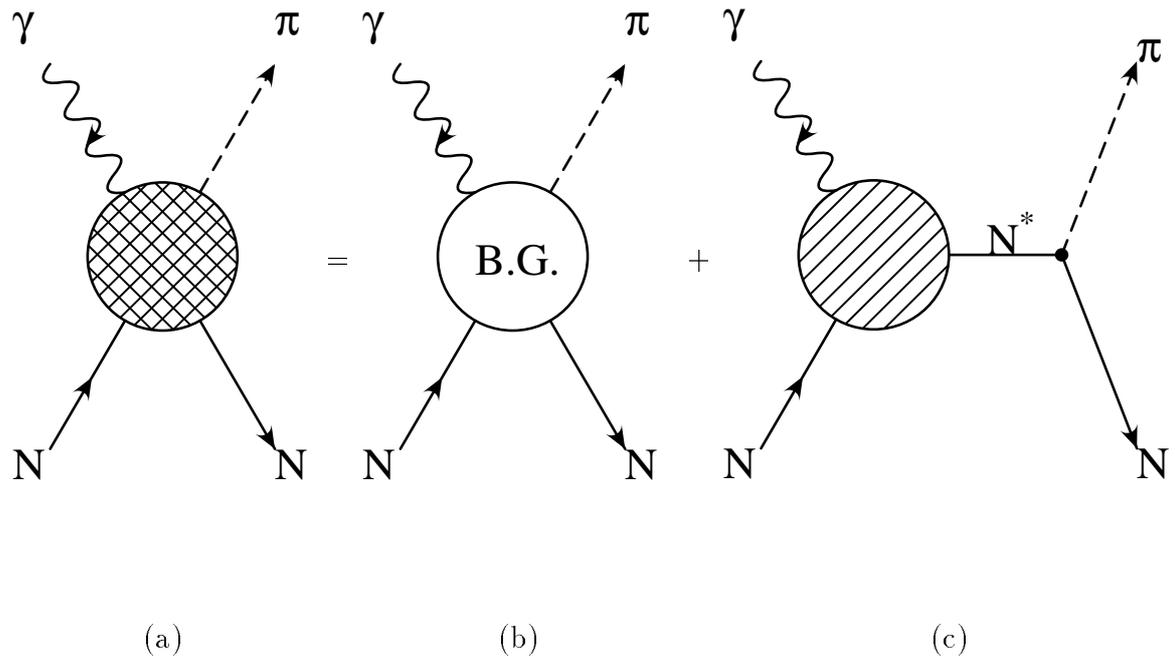
\begin{figure}[H]
\caption{
The $\gamma N \rightarrow \pi N$ reaction in the D$_{13}$ channel
including background and $N^{*}$ production.
B.G. corresponds to the background term.
(a) The full-D$_{13}$ amplitude. 
(b) The Born term for the $\gamma N \rightarrow \pi N$ reaction.
(c) The $N^{*}$ resonant term through the effective $\gamma N N^{*}$ vertex.
}
\label{fig:D13FULL}
\end{figure}
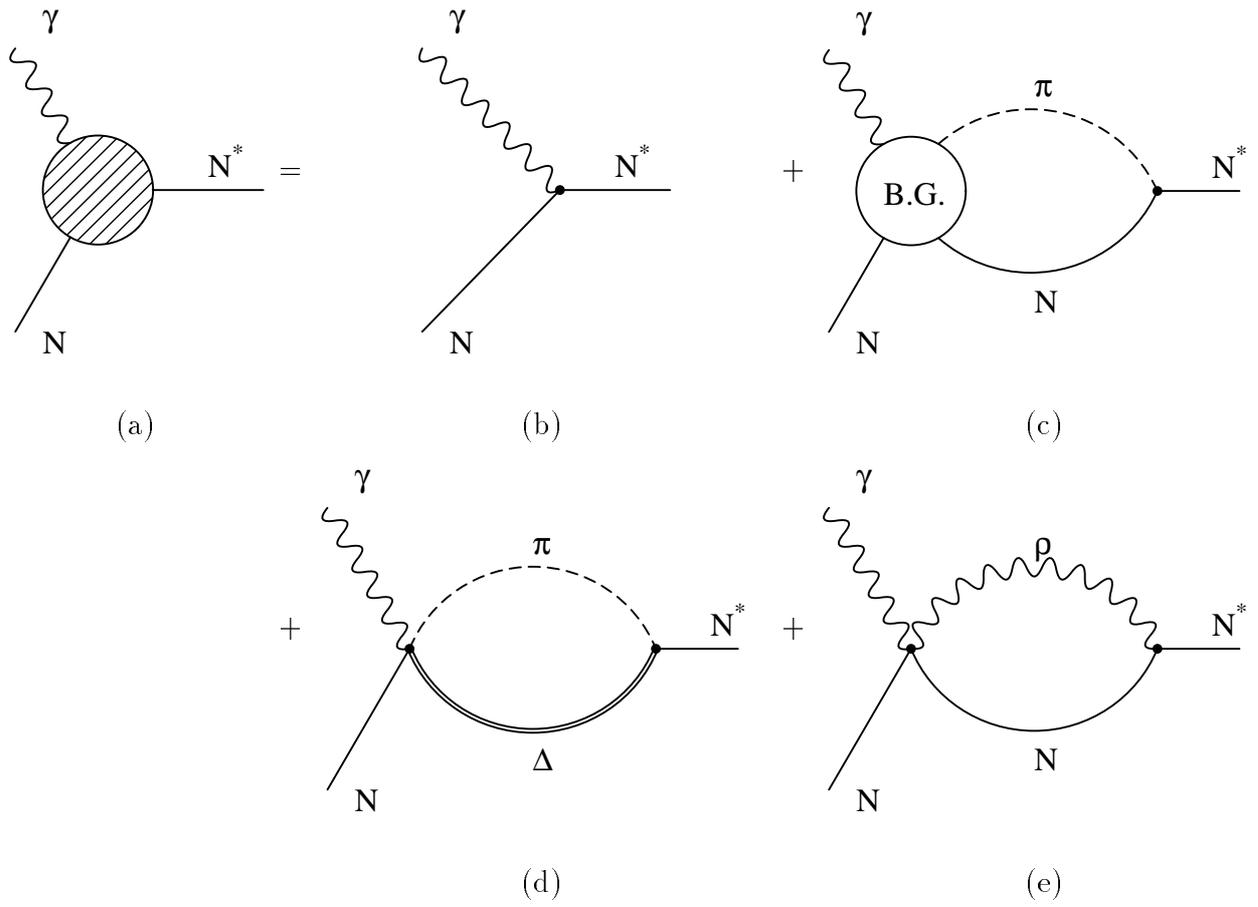
\begin{figure}[H]
\caption{
The effective $\gamma N N^{*}$ vertex.
(a) The effective $\gamma N N^{*}$ vertex.
(b) The bare $\gamma N N^{*}$ vertex.
(c) The vertex correction due to background $\pi N$ production.
(d) The vertex correction due to background $\pi \Delta$ production.
(e) The vertex correction due to background $\rho N$ production.
}
\label{fig:effectivecoupling}
\end{figure}
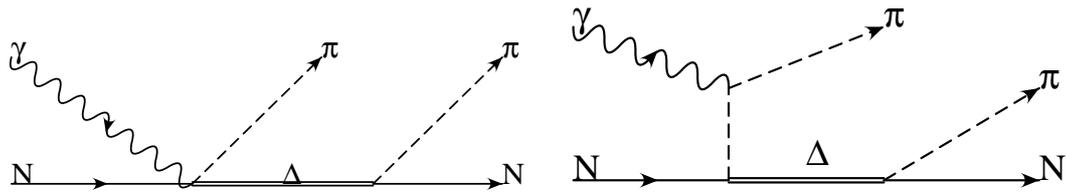
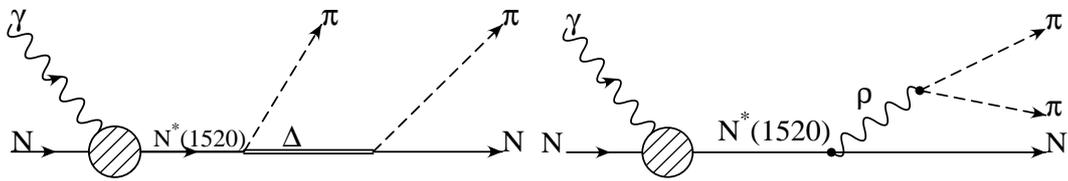
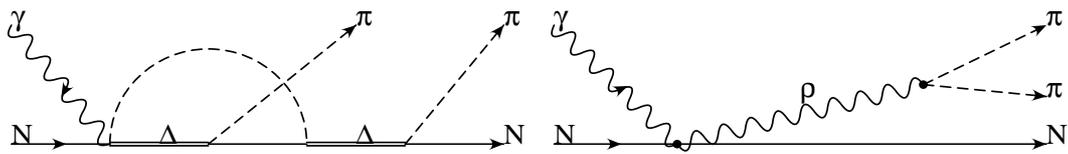
\begin{figure}[H]
\caption{
The diagrams for the $\gamma N \rightarrow \pi \pi N$ reaction.
(a) The $\Delta$ Kroll-Ruderman term.
(b) The $\Delta$ pion-pole term.
(c) The $N^{*} \rightarrow \pi \Delta$ contribution.
(d) The $N^{*} \rightarrow \rho N $ contribution.
(e) The final state rescattering contribution.
(f) The $\rho$-meson Kroll-Ruderman term.
}
\label{fig:2pi-1}
\end{figure}
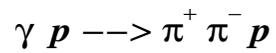
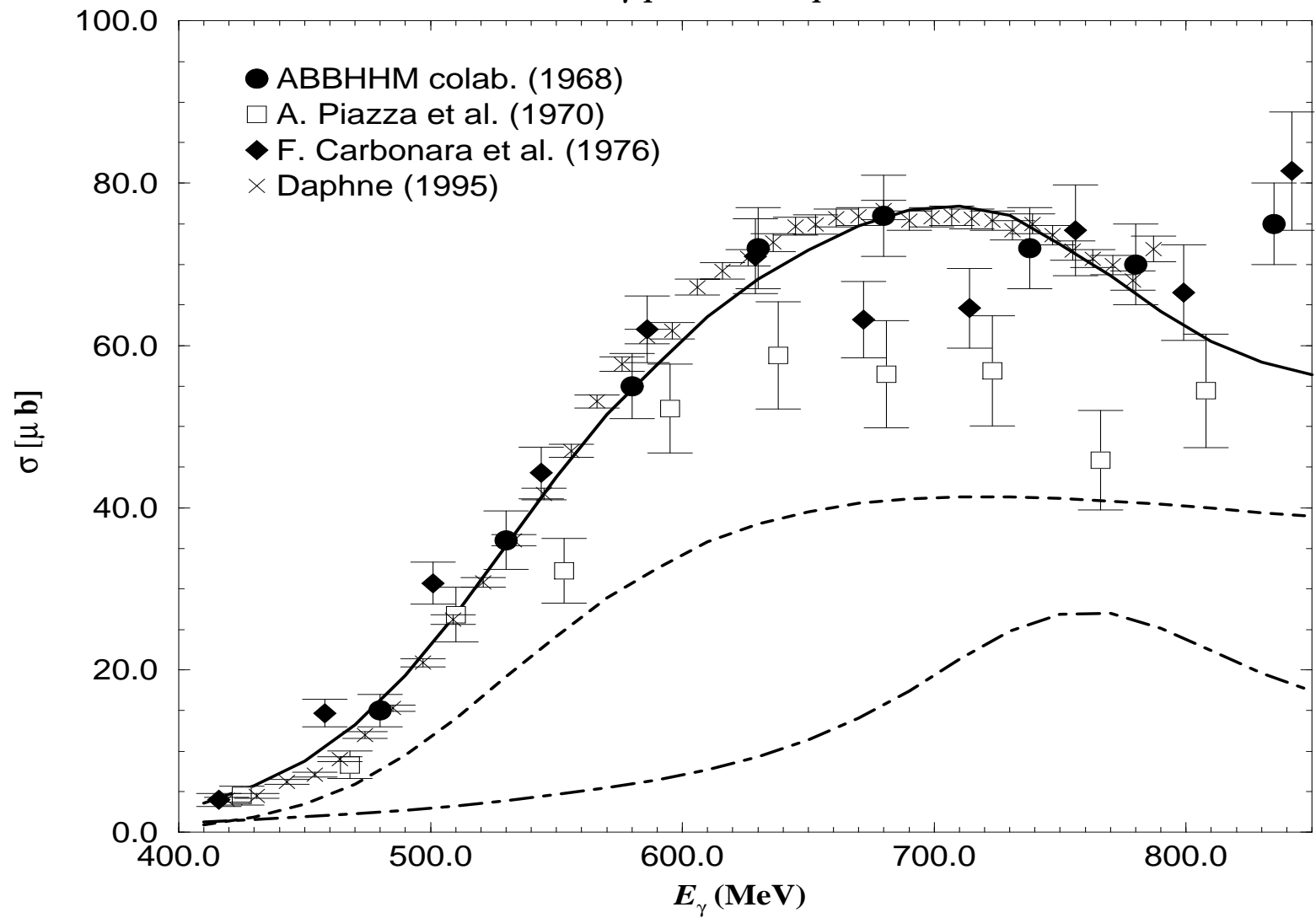
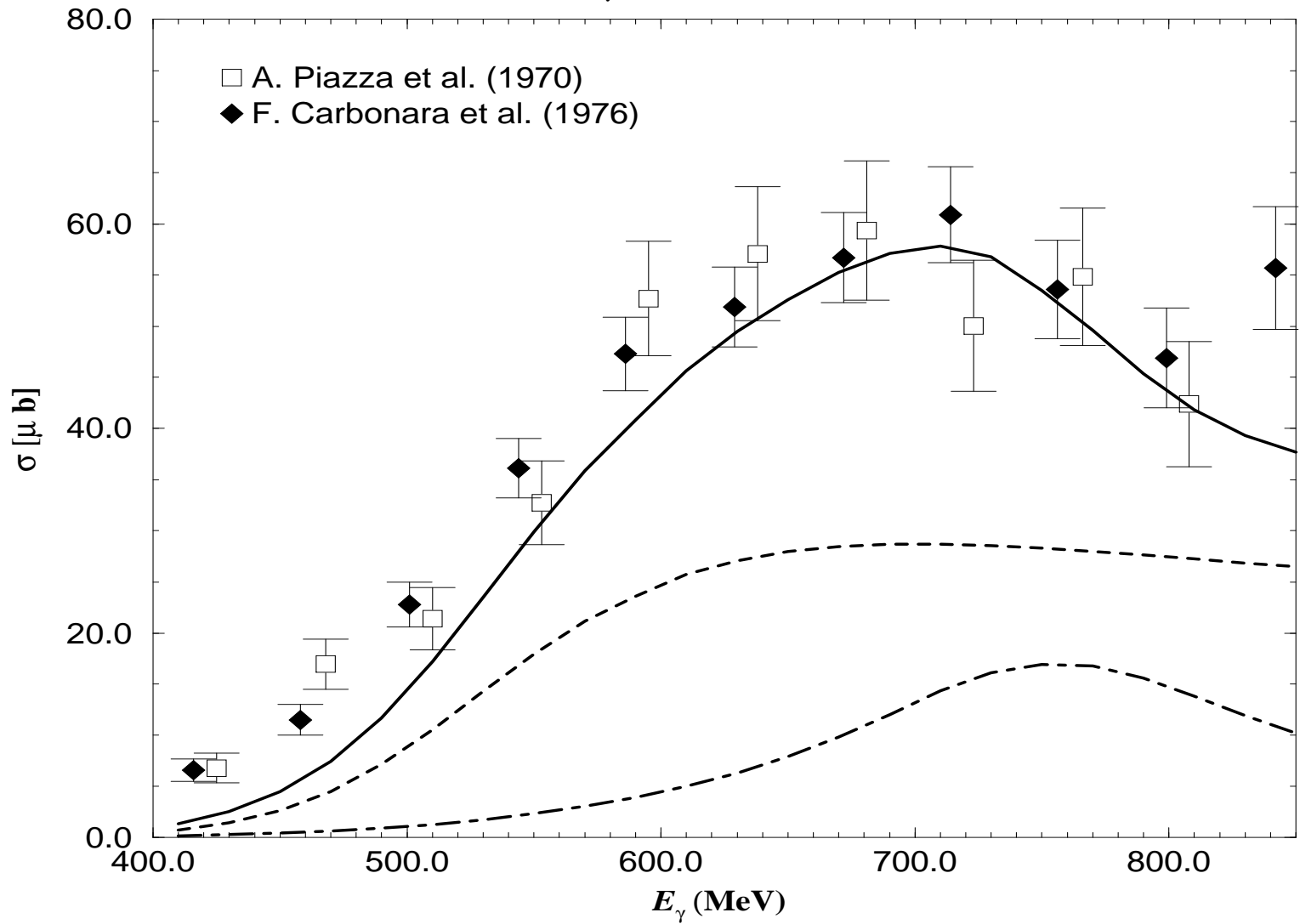
\begin{figure}[H]
\caption{
Total cross sections for (i) the $\gamma p \rightarrow \pi ^{+} \pi ^{-} p$
and (ii) the $\gamma n \rightarrow \pi ^{+} \pi ^{-} n$ reactions.
Solid line corresponds to the
total cross section calculated with the parameter-set(I),
dashed line to the contributions of
the $\Delta$ Kroll-Ruderman and $\Delta$ pion-pole terms
(diagrams (a) and (b) in Fig.\ref{fig:2pi-1})
and
dash-dotted line to the contributions of 
the $N^{*}$ terms (diagrams (c) and (d) in Fig.\ref{fig:2pi-1}).
Experimental data are taken from
Refs. \protect\cite{doublepi-exp1,doublepi-exp2,PIA,CAR}.
}
\label{fig:pmp-pmn}
\end{figure}
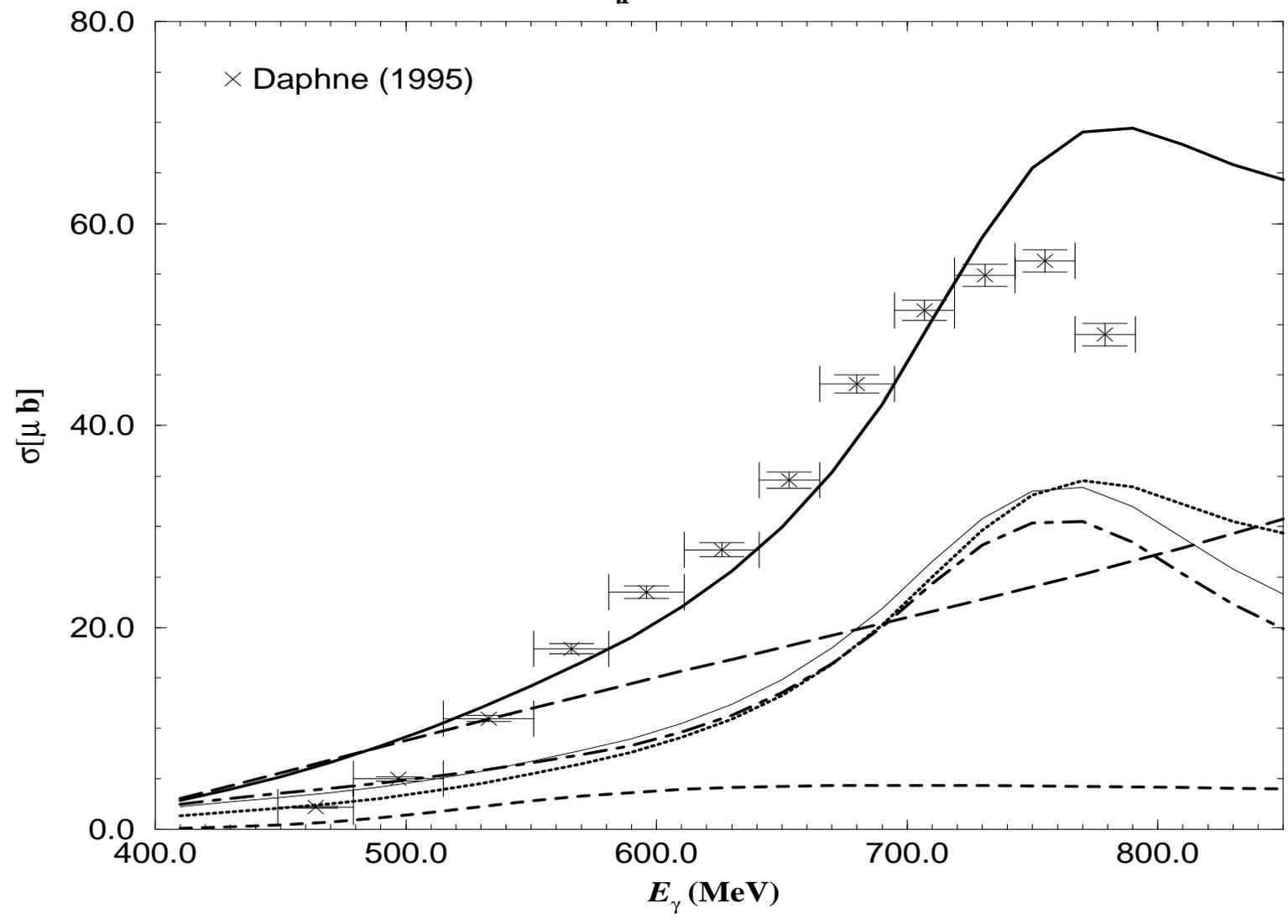
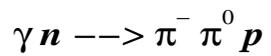
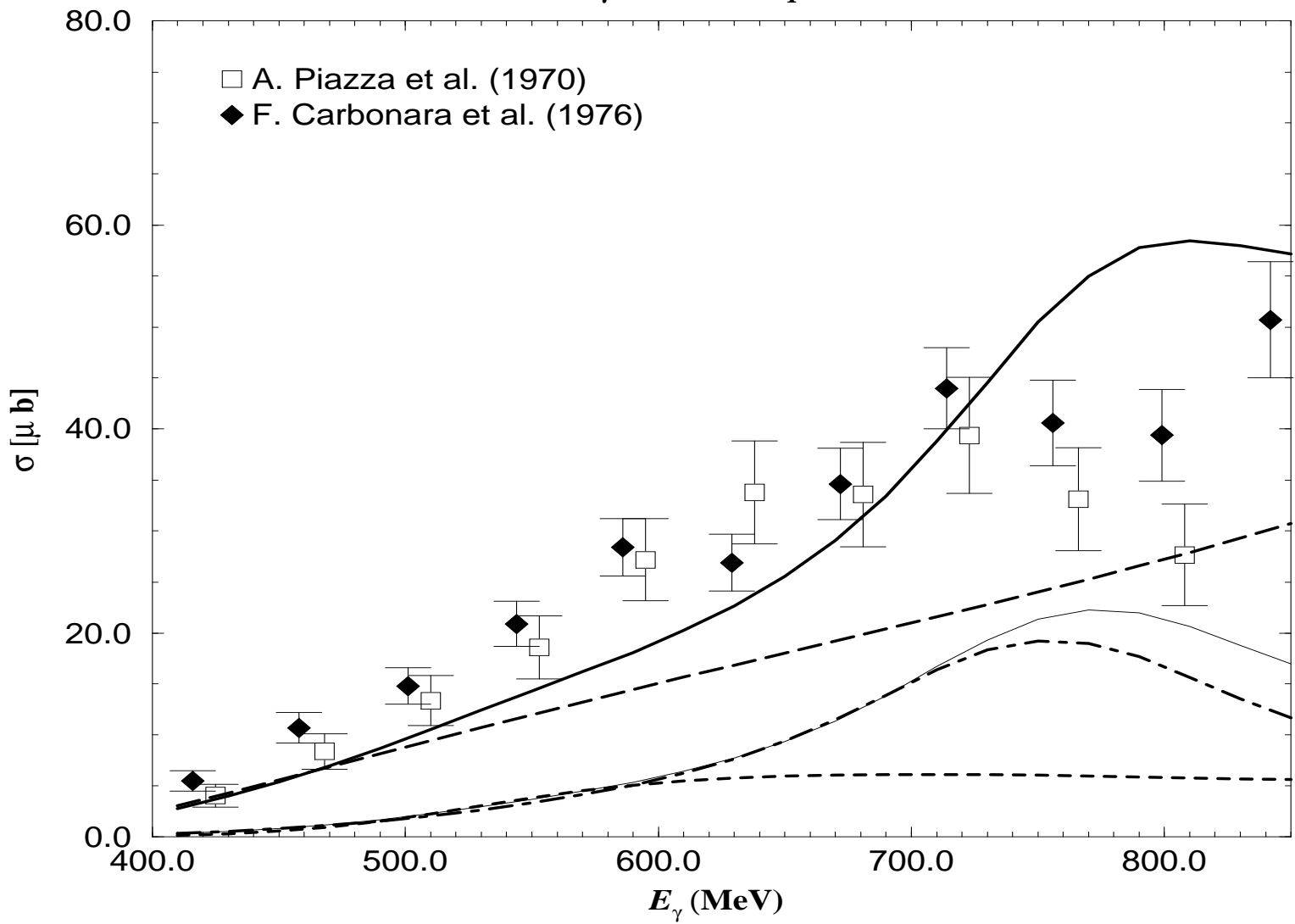
\begin{figure}[H]
\caption{
Total cross sections for
(i) the $\gamma p \rightarrow \pi ^{+} \pi ^{0} n$
and
(ii) the $\gamma n \rightarrow \pi ^{-} \pi ^{0} p$ reactions.
Thin-solid line corresponds to 
the total cross section calculated by using Eq.(\ref{eq:T1}) 
with the parameter-set (I) and
dotted line to the total cross section calculated 
by using Eq.(\ref{eq:T2}) with the parameter-set (I).
Bold-solid line corresponds to the
total cross section calculated by using Eq.(\ref{eq:T2}) with 
the parameter-set (II), 
short-dashed line to the contributions of
the $\Delta$ Kroll-Ruderman and $\Delta$ pion-pole terms
(diagrams (a) and (b) in Fig.\ref{fig:2pi-1}),
dash-dotted line to the contributions of
the $N^{*}$ terms (diagrams (c) and (d) in Fig.\ref{fig:2pi-1})
and long-dashed line to the contribution of
the $\rho$-meson Kroll-Ruderman term (diagram (f) in Fig.\ref{fig:2pi-1})
including the finite-ranged form factor of $\rho \pi \pi$.
Experimental data are taken from Refs. \protect\cite{doublepi-exp2,PIA,CAR}.
}
\label{fig:pzn-mzp}
\end{figure}
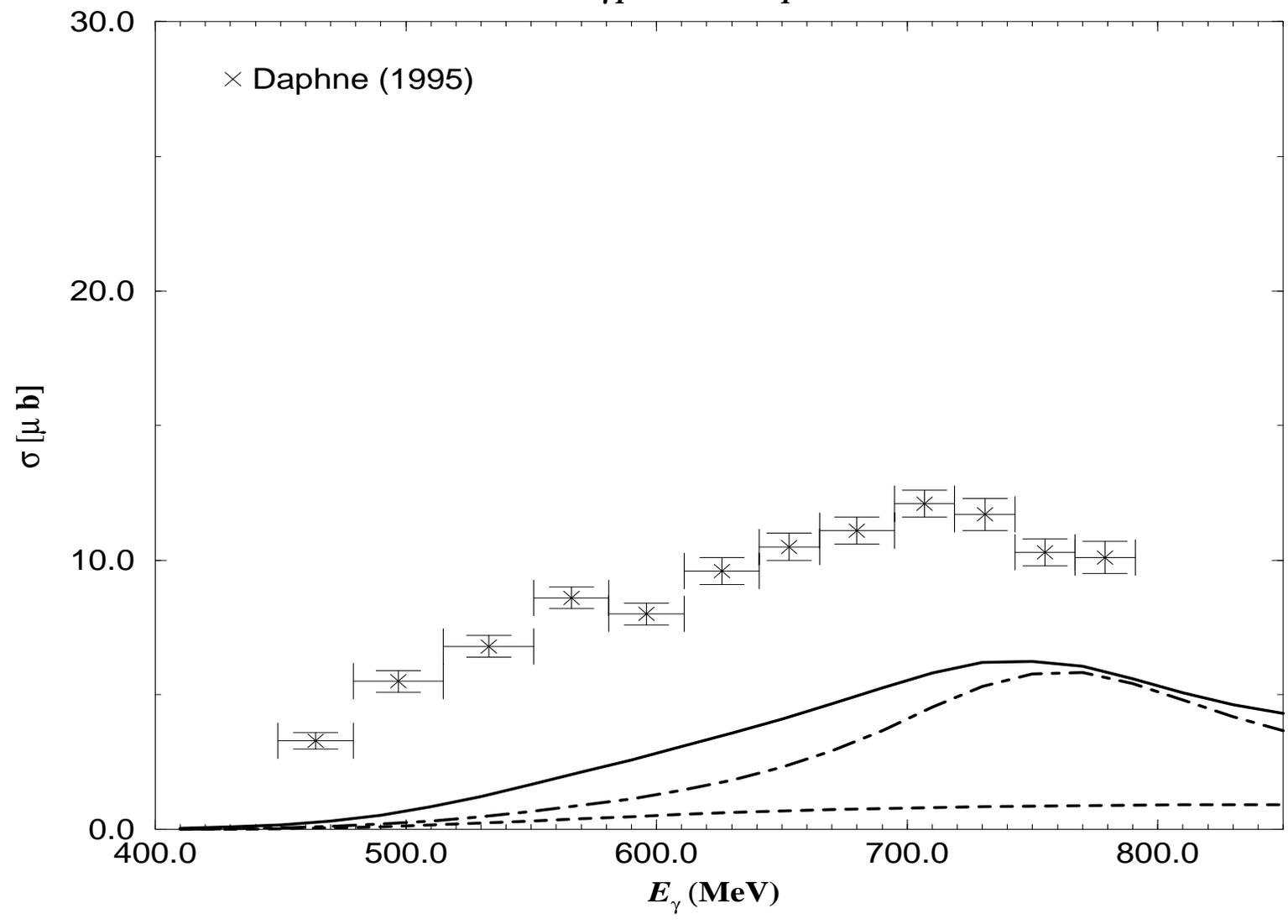
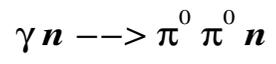
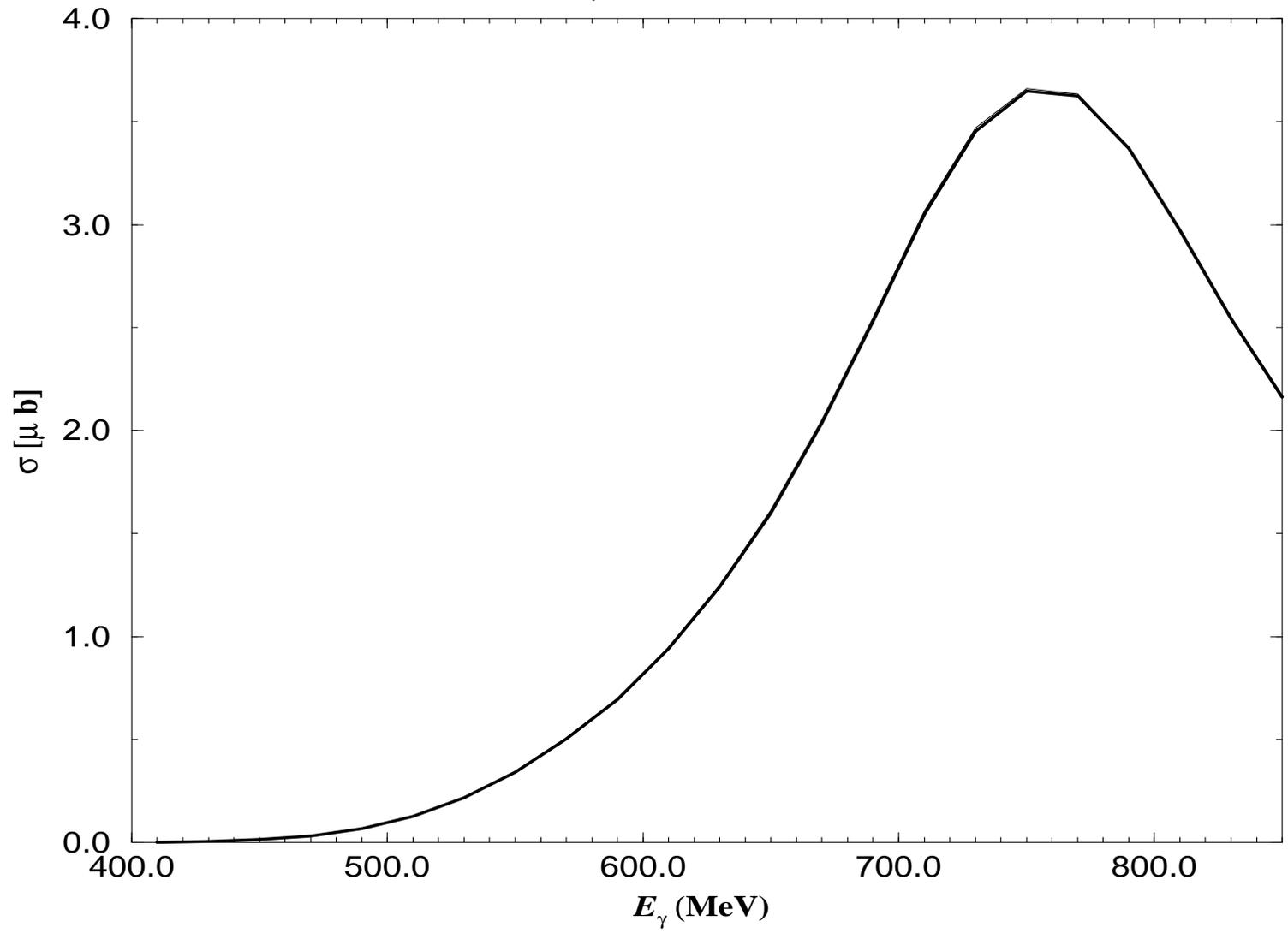
\begin{figure}[H]
\caption{
Total cross sections for
(i) the $\gamma p \rightarrow \pi ^{0} \pi ^{0} p$
and
(ii) the $\gamma n \rightarrow \pi ^{0} \pi ^{0} n$ reactions.
Solid line corresponds to the total cross section
calculated by using Eq.(\ref{eq:T2}) with the parameter-set (II),
dashed line to the contributions of
the $\Delta$ Kroll-Ruderman term (diagram (a) in Fig.\ref{fig:2pi-1}) and
dash-dotted line to the contribution of
the $N^{*}$ term (diagram (c) in Fig.\ref{fig:2pi-1}).
Experimental data is taken from Ref. \protect\cite{doublepi-exp2}.
}
\label{fig:zzp-zzn}
\end{figure}
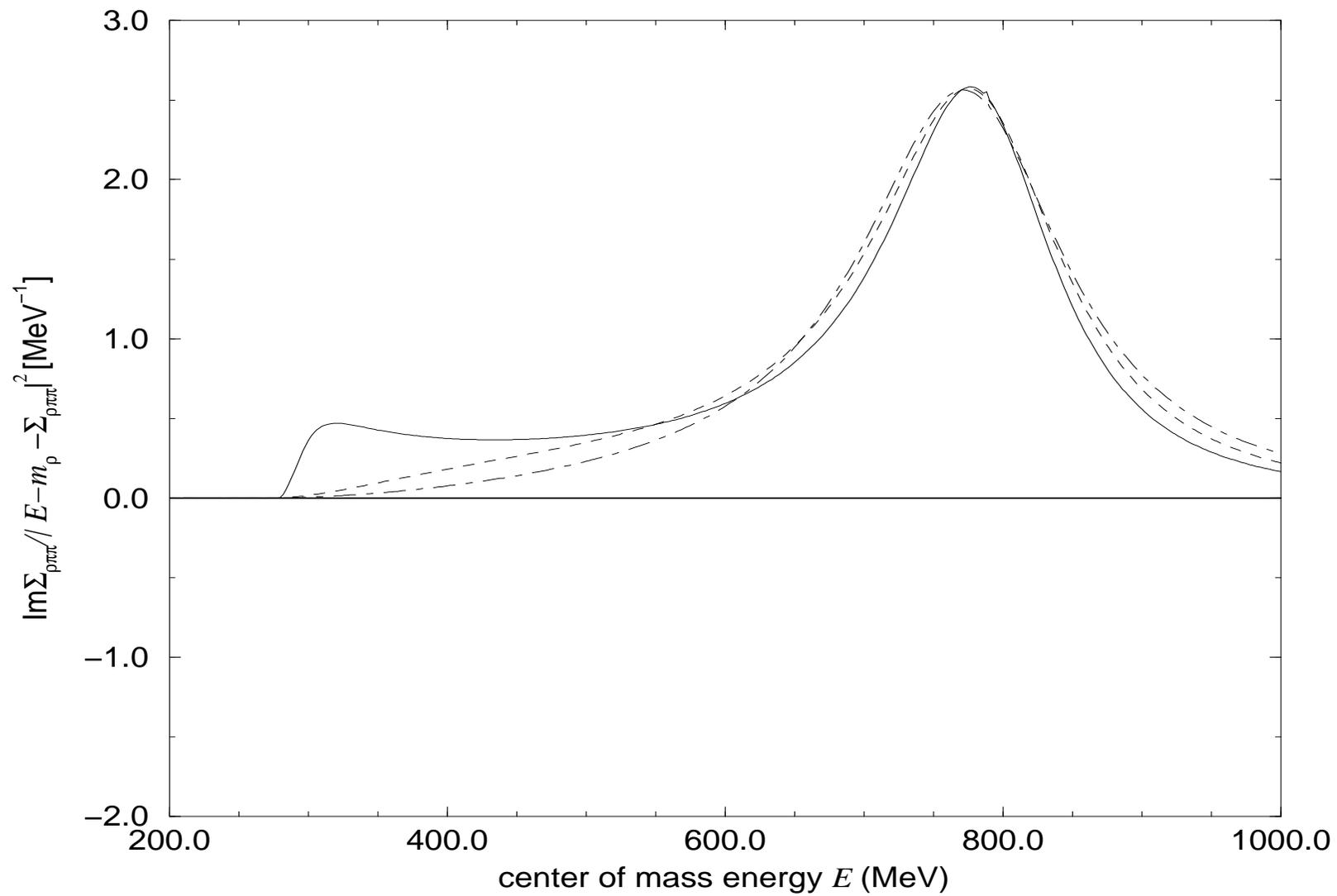
\begin{figure}[H]
\caption{
The 2$\pi$-spectral function as a function of the center-of-mass energy.
Solid line corresponds to $q_{\rho \pi \pi}= 100$ MeV/c,
dashed line to $q_{\rho \pi \pi}= 200$ MeV/c and
dash-dotted line to $q_{\rho \pi \pi}= 300$ MeV/c.
}
\label{fig:pipiamp}
\end{figure}
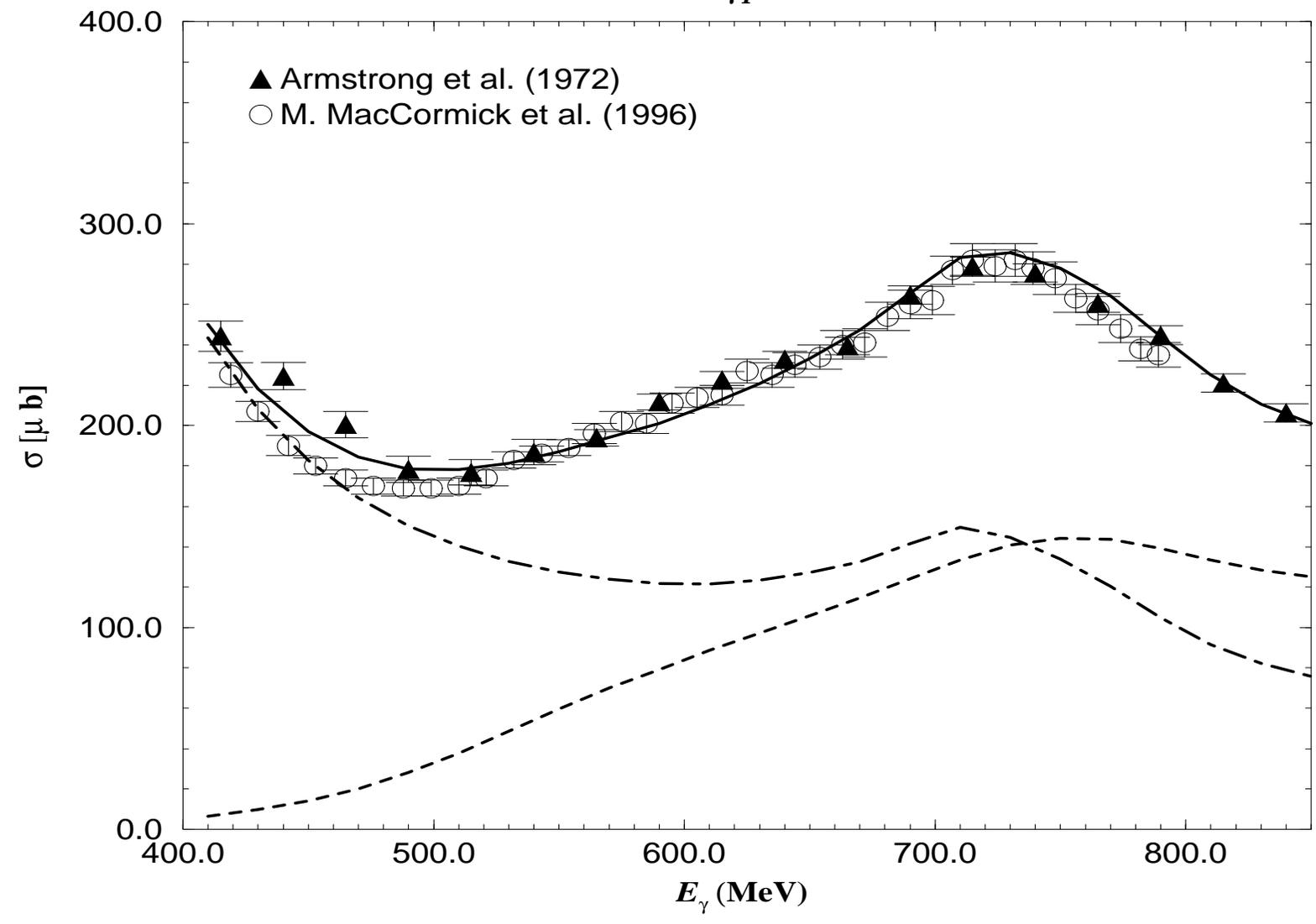
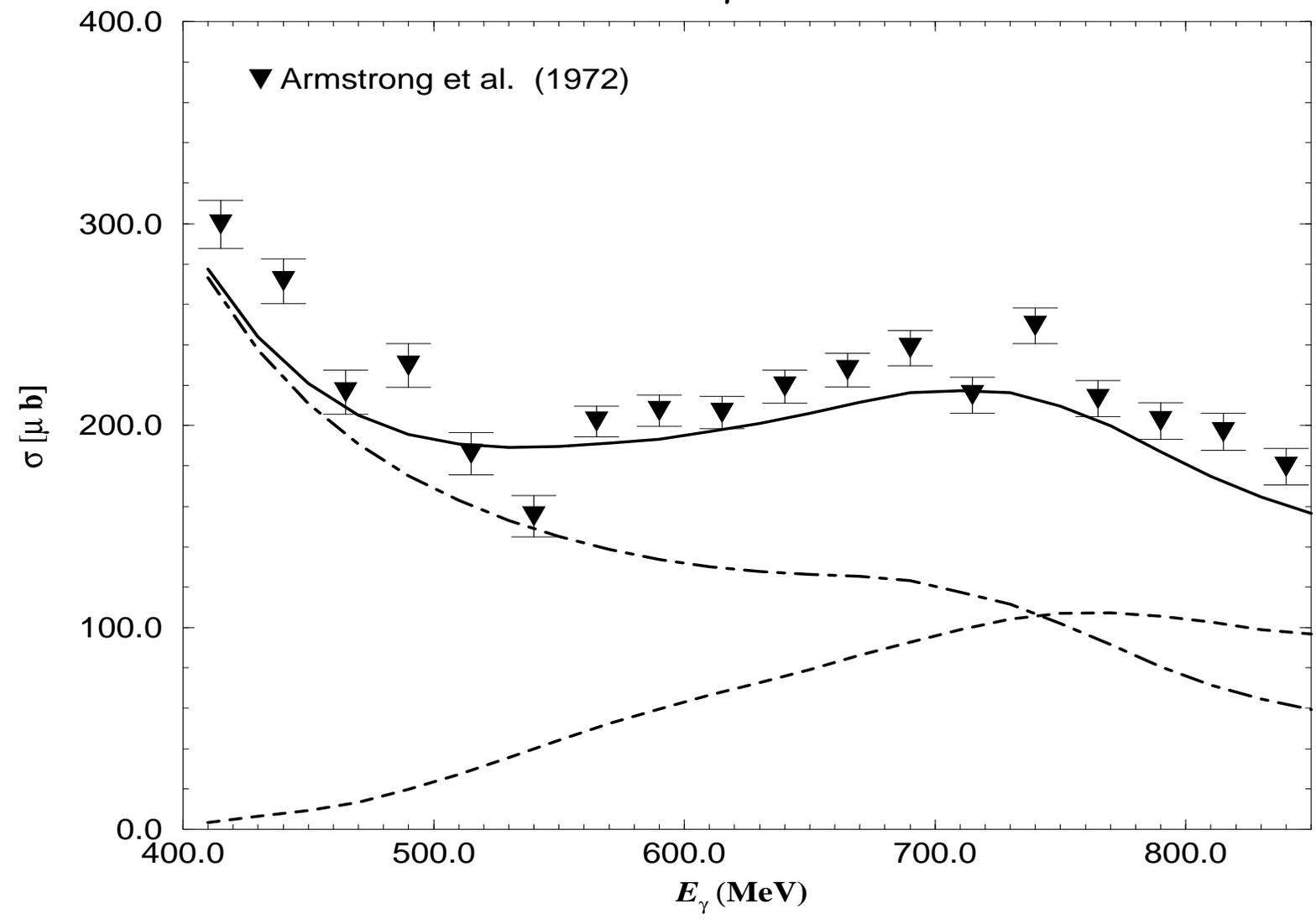
\begin{figure}[H]
\caption{
The total photoabsorption cross section of
(i) the proton target and (ii) the neutron target.
Solid line corresponds to the 
summed cross section of $\gamma N \rightarrow \pi N$ and
$\gamma N \rightarrow \pi \pi N$ which are calculated in our model.
Dashed line corresponds to the contribution of 
$\gamma N \rightarrow \pi N$,
dash-dotted line to the contribution of
$\gamma N \rightarrow \pi \pi N$.
Experimental data are taken from Ref. \protect\cite{Nucl4,Arm-p,Arm-n}.
}
\label{fig:total-p-n}
\end{figure}
\newpage
\onecol

\begin{references}
%
\bibitem{Nucl1}
M. Anghinolfi {\it et al}., Phys. Rev. C {\bf 47} (1993) R992.
%
\bibitem{Nucl2}
N. Bianchi {\it et al}., Phys.Lett. B {\bf 229} (1993) 219.
%
\bibitem{Nucl3}
N. Bianchi {\it et al}., Phys. Lett. B {\bf 309} (1993) 5.
%
\bibitem{Nucl4}
M. MacCormick {\it et al}., Phys. Rev. C {\bf 53} (1996) 41.
%
\bibitem{Nucl5}
N. Bianchi {\it et al}., Phys. Rev. C{\bf 54} (1996) 1688.
%
\bibitem{Phenomenan1}
W.M. Alberico, G. Gervino and A. Lavagno, Phys. Lett. B 321.
(1994) 177
%
\bibitem{Phenomenan2}
L.A. Kondratyuk, M.I. Krivoruchenko, N. Bianchi, E. De Sanctis,
V. Mucchifora, Nucl. Phys. A {\bf 579} (1994) 453.
%
\bibitem{Phenomenan3}
M. Effenberger, A. Hombach, S. Teis and U. Mosel, preprint nucl-th/9607005.
%
\bibitem{singlepi-exp}
see, for example,
C. Betourne {\it et al}., Phys. Rev. {\bf 172} (1968) 11343,
S.D. Ecklund and R.L. Walker, {\it ibid}. {\bf 159} (1967) 1195,
H. De Staebler {\it et al}., {\it ibid}. {\bf 140} (1965) B337.
%
\bibitem{Moorhouse}
R.G. Moorhouse, H. Oberlack and A.H. Rosenfeld,
Phys. Rev. D {\bf 9} (1974) 1.
%
\bibitem{Fujii}
I. Arai and H. Fujii, Nucl. Phys. {\bf B194} (1982) 251.
%
\bibitem{Arndt}
R.A. Arndt, R.L. Workman, Z. Li and L.D. Roper,
Phys. Rev. C {\bf 42} (1990) 1853,
and the Scattering Analysis Interactive Dial-in (SAID) program,
available by TELNET call to clsaid.phys.vt.edu,
or WWW site at http://clsaid.phys.vt.edu.
%
\bibitem{doublepi-exp1}
Aachen-Berlin-Bonn-Hamburg-Heidelberg-M\"unchen collaboration,
Phys. Rev. {\bf 175} (1968) 1669.
%
\bibitem{doublepi-exp2}
A. Braghieri {\it et al}., Phys. Lett. B {\bf 365} (1995) 46 .
%
\bibitem{Oset}
J.A.G. Tejedor, E. Oset, Nucl. Phys. A{\bf 571} (1994) 667,
preprint hep-ph/9506209.
%
\bibitem{Laget}
L.Y. Murphy and J.M. Laget, preprint DAPNIA-SPHN-95-42.
%
\bibitem{PIA}
A. Piazza {\it et al}., Nuovo Cimento {\bf III} (1970) 403.
%
\bibitem{CAR}
F. Carbobara {\it et al}., Nuovo Cimento {\bf 36 A} (1976) 219.
%
\bibitem{Particledata}
Particle Data Group, Phys. Rev. D {\bf 50} (1994).
%
\bibitem{Liu}
R.S. Bhalerao and L.C. Liu, Phys. Rev. Lett. {\bf 54} (1985) 865.
%
\bibitem{Arima}
M. Arima, K. Masutani, R. Seki, Phys. Rev. C {\bf 51} (1995) 285.
%
\bibitem{BetzLee}
M. Betz and T.-S.H. Lee,
Phys. Rev. C {\bf 23} (1981) 375.
%
\bibitem{KMO}
J.H. Koch, E.J. Moniz, and N. Ohtsuka,
Ann. Phys. {\bf 154} (1984) 99.
%
\bibitem{Lee}
S. Nozawa, B. Blankleider and T.-S.H. Lee,
Nucl. Phys. A {\bf 513} (1990) 459.
%
\bibitem{Kumano}
L. Heller, S. Kumano, J.C. Martinez and E.J. Moniz,
Phys. Rev. C {\bf 35} (1987) 718.
%
\bibitem{MC}
S. Kawabata, Comput. Phys. Commun. {\bf 41} (1986) 127.
%
\bibitem{Toki}
E. Oset, H. Toki and W. Weise, Phys. Rep. {\bf 83} No.4 (1982) 281.
%
\bibitem{Arm-p}
T.A. Armstrong, {\it et al}., Phys. Rev. D {\bf 5} (1972) 1640.
%
\bibitem{Arm-n}
T.A. Armstrong, {\it et al}., Nucl. Phys. B {\bf 41} (1972) 445.
%
\bibitem{future}
K. Ochi, M. Hirata and T. Takaki, in preparation.
\end{references}
\end{document}